\documentstyle[11pt,epsf]{article}
\newtheorem{theorem}{Theorem}
\newtheorem{corollary}{Corollary}

\newcommand{\prm}{m^\prime}
\newcommand{\tm}{\tilde{m}}
\newcommand {\dfn} {\stackrel{\Delta} {=}}
\newcommand {\exe} {\stackrel{\cdot} {=}}
\newcommand {\lexe} {\stackrel{\cdot} {\le}}
\newcommand{\eqa}{\stackrel{\mbox{(a)}}{=}}
\newcommand{\eqb}{\stackrel{\mbox{(b)}}{=}}

\newcommand {\reals} {{\rm I\!R}}

\newcommand {\bu} {\mbox{\boldmath $u$}}
\newcommand {\bv} {\mbox{\boldmath $v$}}

\newcommand {\bx} {\mbox{\boldmath $x$}}
\newcommand {\by} {\mbox{\boldmath $y$}}
\newcommand {\bz} {\mbox{\boldmath $z$}}

\newcommand {\bE} {\mbox{\boldmath $E$}}

\newcommand {\tsigma} {\tilde{\sigma}}
\newcommand {\hsigma} {\hat{\sigma}}

\newcommand {\hC} {\hat{C}}
\newcommand {\hI} {\hat{I}}

\newcommand {\bX} {\mbox{\boldmath $X$}}
\newcommand {\bY} {\mbox{\boldmath $Y$}}
\newcommand {\bZ} {\mbox{\boldmath $Z$}}
\newcommand{\calA}{{\cal A}}

\newcommand{\calC}{{\cal C}}

\newcommand{\calE}{{\cal E}}
\newcommand{\calF}{{\cal F}}

\newcommand{\calH}{{\cal H}}
\newcommand{\calI}{{\cal I}}

\newcommand{\calN}{{\cal N}}

\newcommand{\calT}{{\cal T}}

\newcommand{\calZ}{{\cal Z}}

\begin{document}
\thispagestyle{empty}
\title{Error Exponents of Typical Random Codes\\ for the Colored Gaussian
Channel
\thanks{This research was supported by Israel Science Foundation (ISF) grant
no.\ 137/18.}}
\author{Neri Merhav}
\date{}
\maketitle

\begin{center}
The Andrew \& Erna Viterbi Faculty of Electrical Engineering\\
Technion - Israel Institute of Technology \\
Technion City, Haifa 32000, ISRAEL \\
E--mail: {\tt merhav@ee.technion.ac.il}\\
\end{center}
\vspace{1.5\baselineskip}
\setlength{\baselineskip}{1.5\baselineskip}

\begin{abstract}
The error exponent of the typical random code is defined as the asymptotic normalized
expectation of the logarithm of the probability of error, as opposed
to the traditional definition of the random coding exponent as the
normalized logarithm of the expectation of the probability of error with
respect to a given ensemble of codes. For a certain ensemble of independent codewords,
with a given power spectrum, and a generalized stochastic mismatched decoder, 
we characterize the error exponent the typical
random codes (TRC) for the colored Gaussian channel, 
with emphasis on the range of low rates, where the TRC error exponent differs
in value from the ordinary random coding error exponent. The error exponent
formula, which is exponentially tight at some range of low rates, is presented
as the maximum of a certain function with respect to one parameter only
(in the spirit of Gallager's formulas) in the case of matched decoding, and
two parameters in the case of mismatched decoding.
Several aspects of the main results are discussed. These include: general
properties, a parametric
representation, critical rates,
phase transitions, optimal input spectrum (water pouring), 
and comparison to the random coding
exponent.\\

\end{abstract}

\newpage
\section*{I.\ Introduction}

Inspired by the brief article of Barg and Forney \cite{BF02}, in a recent work \cite{trc},
the error exponent of the {\it typical random code} (TRC) 
for a general discrete memoryless channel (DMC) was studied. The error
exponent of the TRC was defined as the asymptotic
normalized {\it expectation of the logarithm} of the probability of error, as opposed
to the traditional definition of the random coding exponent as the
normalized {\it logarithm of the expectation} of the probability of error with
respect to the same ensemble of codes. The study of error exponents for TRCs
was motivated in \cite[Introduction]{trc} in several ways: (i) due to Jensen's inequality, 
it is never worse than the random coding error exponent. 
(ii) In relation to (i), it is a less pessimistic
performance metric compared to the ordinary random coding exponent, especially
at low rates, as it does not suffer from the problem that poor codes dominate
the average error probability. (iii) Given that a certain concentration
property holds, it is more relevant as a performance
metric, as the code is normally assumed to be randomly selected just once, and
then it is used repeatedly. (iv) It captures correctly the behavior of 
random--like codes
\cite{Battail95},
which are well known to be very good codes.

In \cite{trc}, 
an exact formula for the error exponent function of the
TRC was derived for a general discrete memoryless channel (DMC) under the
ensemble of fixed composition codes and a class of generalized likelihood
decoders (GLD's) \cite{gld}, \cite{SMF15}, \cite{YAG13}, 
namely, stochastic decoders that randomly select the decoded message according
to a posterior distribution (given the channel output) with a
certain structure. The class of GLD's considered in \cite{trc} covers many
decoders of theoretical and practical interest as 
special cases, including deterministic metric
decoders, like the maximum likelihood (ML) decoder, the maximum mutual
information (MMI) decoder, $\alpha$--decoders \cite{CK81} and mismatched decoders.

While the analysis in \cite{trc} is heavily based on the method of types
\cite{CKbook}, and
hence applicable, in principle, to the finite--alphabet case only, here we consider
the continuous--alphabet case, and more precisely, the Gaussian case.
In particular, we derive
a formula for the error exponent of the TRC
for the additive Gaussian channel. We begin from the additive white Gaussian
noise (AWGN) channel, and then extend the scope to the colored Gaussian
channel with a given channel input spectrum 
and noise spectrum. While we present the results for the
discrete--time case, a minor modification of the error exponent formula allows to pass to the
continuous--time case as well. 

It is perhaps surprising that although the
finite--alphabet formula of the TRC exponent \cite{trc} is not trivial to work with (as
it involves a rather computationally 
heavy optimization of certain functionals of probability
distributions), in the Gaussian noise case considered here (and even when the
noise is colored), the
situation is considerably better in that sense. In particular, the
resulting TRC error exponent, which is provably exact at least at some range of low
rates,\footnote{In fact, the TRC exponent function is interesting to study
primarily at low
rates, as beyond a certain rate, it coincides with the ordinary random coding
error exponent anyway.} involves optimization 
over one parameter only in the case of a GLD with the matched
(ML) decoding metric, and two parameters for a general (mismatched) GLD. 
Finally, we present and discuss a few byproducts of our main result. These include:
a parametric representation,
the zero--rate TRC exponent, the range of guaranteed tightness of our bound, the
rate at which the TRC exponent meets the random coding exponent, and the optimal
channel input spectrum (water--pouring).

A few words on the history of related work are in order. 

In the context of bounds on the reliability function of the Gaussian channel,
the first random coding bound for the discrete--time case, as well
as a sphere--packing lower bound, are due to Shannon \cite{Shannon59}. 
His work was followed by numerical evaluations due to Slepian \cite{Slepian63}. Ebert
\cite{Ebert66} has derived (random--coding and expurgated) upper bounds and
(sphere--packing and straight--line) lower bounds on the error probability of 
the parallel additive Gaussian channel, which in the long block limit, are
applicable to the colored Gaussian channel by using the eigenvalue distribution
theorem \cite{GS58} (see also \cite{Gray06}). 
The lower bounds are based on the work of Shannon, Gallager and
Berlekamp \cite{SGB67}, which appeared a year later.
Ebert's results appeared also in Gallager's book
\cite[Chapters 7, 8]{Gallager68}. Viterbi \cite{Viterbi69} studied decoders
with generalized decision regions for the Gaussian channel as well as other
very noisy channels. During the years that have passed ever since these early
works were published, a vast amount of work on improved bounds has been
carried out (see, e.g., \cite{SS06} for a very good survey), but most of it is
of lesser direct relevance to the topic of this paper. 

In the context of error exponents of
TRC's, on the other hand, much less work has been done before. Prior to \cite{trc},
as mentioned earlier, 
Barg and Forney \cite{BF02} have derived, among other things, the error exponent of the TRC 
for the binary symmetric channel (BSC) under i.i.d.\ binary symmetric random
coding. Nazari \cite{Nazari11} and Nazari, Anastasopoulos and Pradhan 
\cite{NAP14} derived upper and lower bounds
on the TRC exponent for a general DMC under the $\alpha$--decoder of
\cite{CK81}. Beyond that, there has been some work in the statistical--physics
literature, where TRC error exponents were analyzed for special classes of
codes (like LDPC and Turbo codes) using the replica method and the cavity
method -- see, e.g., \cite{Kabashima08}, \cite{MR06},
\cite{SMKS03}.

The outline of the remaining part of this paper is as follows. In Section II,
we establish notation conventions (Subsection II.A), 
present the setup (Subsection II.B), and provide some background (Subsection
II.C). In Section III, we provide the main results concerning the TRC
exponent and discuss them, 
first for the case of the AWGN channel (Subsection III.A), and then for parallel
channels and the colored Gaussian channel (Subsection III.B).
In Section IV, we address the question of the optimal input spectrum for the
TRC exponent. Finally, in Section V, we outline the proofs of the main
results.

\section*{II.\ Notation Conventions, Setup and Background}

\subsection*{A.\ Notation Conventions}

Throughout the paper, random variables will be denoted by capital
letters and specific values they may take will be denoted by the
corresponding lower case letters. Random
vectors and their realizations will be denoted,
respectively, by capital letters and the corresponding lower case letters,
both in the bold face font.
For example, the random vector $\bX=(X_0,X_1,\ldots,X_{n-1})$, ($n$ --
positive integer) may take a specific vector value $\bx=(x_0,x_1,\ldots,x_{n-1})$
in $\reals^n$. When used in the linear--algebraic context, these vectors
should be thought of as column vectors, and so, when they appear with
superscript $T$, they will be transformed into row vectors by transposition.
Thus, $\bx^T\by$ is understood as the inner product of $\bx$ and $\by$.
Probability density functions (PDFs) of sources and 
channels will be denoted by the letters $P$ and $Q$.
The probability of an event $\calE$ will be denoted by $\mbox{Pr}\{\calE\}$,
and the expectation
operator will be
denoted by $\bE\{\cdot\}$. 

For two positive sequences $a_n$ and $b_n$, the notation $a_n\exe b_n$ will
stand for equality in the exponential scale, that is,
$\lim_{n\to\infty}\frac{1}{n}\log \frac{a_n}{b_n}=0$. Similarly,
$a_n\lexe b_n$ means that
$\limsup_{n\to\infty}\frac{1}{n}\log \frac{a_n}{b_n}\le 0$, and so on.
Accordingly, $a_n\exe e^{-n\infty}$ will mean that $a_n$ vanishes
in a super--exponential rate.
The indicator function
of an event $\calE$ will be denoted by $\calI\{E\}$. The notation $[x]_+$
will stand for $\max\{0,x\}$. The cardinality of a finite set $\calA$ will be
denoted by $|\calA|$.

\subsection*{B.\ Setup}

We first describe the communication system model in the simple case of the
AWGN channel, and then extend the scope to parallel Gaussian channels with
application to the colored Gaussian channel.

\subsubsection*{1.\ The AWGN Channel}

Consider the discrete--time AWGN channel,
\begin{equation}
Y_i=X_i+Z_i,~~~~i=0,1,\ldots,n-1
\end{equation}
where $\bX=(X_0,\ldots,X_{n-1})$ is a random channel input vector,
$\bZ=(Z_0,\ldots,Z_{n-1})$ is a zero--mean Gaussian vector with
covariance matrix $\sigma^2I$, $I$ being the $n\times n$ identity matrix,
and $\bZ$ is statistically independent of $\bX$. 
In the sequel, we denote the conditional
probability density function (PDF) associated with this channel by
$P(\by|\bx)$, that is,
\begin{equation}
P(\by|\bx)=(2\pi\sigma^2)^{-n/2}\exp\left\{-\frac{1}{2\sigma^2}\sum_{i=0}^{n-1}
(y_i-x_i)^2\right\}.
\end{equation}
It is assumed that $\bX$ is
uniformly distributed across a codebook,
$\calC_n=\{\bx[0],\bx[1],\ldots,\bx[M-1]\}$, $\bx[m]\in\reals^n$,
$m=0,1,\ldots,M-1$, with $M=e^{nR}$, $R$ being the coding
rate in nats per channel use. 

We consider the random selection of the codebook $\calC_n$, where all codewords
are drawn independently under the PDF,
\begin{equation}
\label{randcodedis}
Q(\bx)=\left\{\begin{array}{ll}
\frac{1}{\mbox{Surf}(\sqrt{nP})} & \|\bx\|^2=nP\\
0 & \mbox{elsewhere}\end{array}\right.
\end{equation}
where $\mbox{Surf}(r)$ is the surface area of the $n$--dimensional Euclidean
sphere of radius $r$, and $P$ is the transmission power.

Once the codebook $\calC_n$ has been drawn, it is revealed to both the encoder
and decoder.
We consider the stochastic likelihood decoder, which randomly selects
the decoded message index $\hat{m}$ according to the generalized posterior,
\begin{equation}
\label{posterior1}
P_\beta(\hat{m}=m|\by)=\frac{P^{\beta}(\by|\bx[m])}
{\sum_{m^\prime=0}^{M-1}P^{\beta}(\by|\bx[m^\prime]}=
\frac{\exp\{\beta\bx^T[m]\cdot\by/\sigma^2\}}
{\sum_{m^\prime=0}^{M-1}\exp\{\beta\bx^T[m^\prime]\cdot\by/\sigma^2\}},
\end{equation}
where $\beta> 0$ is a design parameter, which we select freely, and where 
the second equality stems from the assumption that all codewords have
the same norm (energy), as is evidenced in (\ref{randcodedis}).
The motivation for the GLD (\ref{posterior1}) was discussed extensively in
earlier works, such \cite{gld} and \cite{trc} as well as references therein.

The probability of error, for a given code $\calC_n$, is defined as
\begin{equation}
P_{\mbox{\tiny e}}(\calC_n)=\frac{1}{M}\sum_{m=0}^{M-1}\sum_{m^\prime\ne
m}P(\by|\bx_m)\cdot P_\beta(\hat{m}=m^\prime|\by).
\end{equation}

As in \cite{trc}, we define the error exponent of the TRC as
\begin{equation}
E_{\mbox{\tiny trc}}(R)=\lim_{n\to\infty}\left[-\frac{\bE\{\ln P_{\mbox{\tiny
e}}(\calC_n)\}}{n}\right],
\end{equation}
where the expectation is w.r.t.\ the randomness of $\calC_n$, and it is
assumed that
the limit exists. As can be noted, this differs from the
traditional random coding error exponent,
\begin{equation}
E_{\mbox{\tiny r}}(R)=\lim_{n\to\infty}\left[-\frac{\ln[\bE\{P_{\mbox{\tiny
e}}(\calC_n)\}]}{n}\right].
\end{equation}
Our first goal is to derive a single--letter
expression for $E_{\mbox{\tiny trc}}(R)$.

\subsubsection*{2.\ Parallel Additive Gaussian Channels and the Colored Gaussian Channel}

The model of parallel additive Gaussian channels is defined similarly as
in Subsection II.B.1, except that here the various Gaussian noise components
$Z_i$, $i=0,1,\ldots,n-1$, may
have different variances, $\sigma_{i,n}^2$, $i=0,1,\ldots,n-1$, respectively.
In other words, the covariance matrix of $\bZ$ is
$\mbox{diag}\{\sigma_{0,n}^2,\sigma_{1,n}^2,\ldots,\sigma_{n-1,n}^2\}$.
We assume that $\{\sigma_{i,n}^2,~i=0,1,\ldots,n-1\}$ 
obey an asymptotic regime where there exists some function $S_Z(e^{j\omega})$,
$j=\sqrt{-1}$, $\omega\in[-\pi,\pi)$, such that
\begin{equation}
\label{evdt}
\lim_{n\to\infty}\frac{1}{n}\sum_{i=0}^{n-1}G(\sigma_{i,n}^2)=
\frac{1}{2\pi}\int_{0}^{2\pi}G[S_Z(e^{j\omega})]\mbox{d}\omega
\end{equation}
for any continuous function $G:\reals^+\to\reals$.
The underlying motivation of this is the eigenvalue distribution theorem \cite{GS58} (see
also \cite{Gray06}), as will be described in Subsection II.C.
Accordingly, the conditional
PDF associated with this channel is given by
\begin{equation}
P(\by|\bx)=\prod_{i=0}^{n-1}(2\pi\sigma_{i,n}^2)^{-1/2}
\exp\left\{-\frac{1}{2\sigma_{i,n}^2}
(y_i-x_i)^2\right\}.
\end{equation}
The structure of the code will now be slightly more involved than in
Subsection II.B.1. Here we consider a codebook $\calC_N$ of size
$M=e^{NR}$, with block length $N=n\ell$, 
$n$ being the number of parallel channels as before and $\ell$
is another positive integer. Specifically, we subdivide each codeword $\bx[m]\in\calC_N$ 
of length $N$ into $n$
non--overlapping segments, each of length $\ell$, i.e.,
$\bx[m]=(\bx_0[m],\bx_1[m],\ldots,\bx_{n-1}[m])$, 
where $\bx_i[m]=(x_{i\ell}[m],x_{i\ell+1}[m],\ldots,x_{(i+1)\ell-1}[m])$, $i=0,1,\ldots,n-1$.
For every $i$, the segment $\bx_i[m]$ has norm $\|\bx_i[m]\|^2=\ell P_{i,n}$,
and this segment is fed into the $i$--th
channel component whose noise variance is $\sigma_{i,n}^2$. We assume that 
$\sum_{i=0}^{n-1} P_{i,n}\le nP$, where $P$ is the overall power constraint.
As is the case with $\{\sigma_{i,n}^2\}$, asymptotic convergence of 
the error exponent in the large $n$ limit could be expected only if there
would be a well defined asymptotic regime concerning the behavior of 
$\{P_{i,n},~i=0,1,\ldots,n-1\}$ as $n\to\infty$. To this end, we will assume
that $\{P_{i,n}\}$ are ``samples'' of a certain function, $S_X(e^{j\omega})$,
of a continuous, real--valued variable, $\omega\in[0,2\pi)$, that is,
\begin{equation}
\label{psampled}
P_{i,n}=S_X(e^{j2\pi i/n}),~~~~~~~~ i=0,1,\ldots,n-1. 
\end{equation}
We consider the random selection of the codebook $\calC_N$, where all
codewords are drawn independently under the PDF,
\begin{equation}
Q(\bx)=\prod_{i=0}^{n-1} Q_i(\bx_i),
\end{equation}
where
\begin{equation}
\label{randcodedis1}
Q_i(\bx_i)=\left\{\begin{array}{ll}
\frac{1}{\mbox{Surf}(\sqrt{\ell P_{i,n}})} & \|\bx_i\|^2=\ell P_{i,n}\\
0 & \mbox{elsewhere}\end{array}\right.
\end{equation}
Once the codebook $\calC_N$ has been drawn, it is revealed to both the encoder
and decoder.
We consider the stochastic likelihood decoder, which randomly selects
the decoded codeword $\bx\in\calC_N$ according to the generalized posterior,
\begin{equation}
\label{posterior2}
P_\beta(\hat{m}=m|\by)=
\frac{\exp\left\{\beta\sum_{i=0}^{n-1}\bx_i^T[m]\cdot\by_i/\tsigma_{i,n}^2\right\}}
{\sum_{m^\prime=0}^{M-1}
\exp\left\{\beta\sum_{i=0}^{n-1}\bx_i^T[m^\prime]\cdot\by_i/\tsigma_{i,n}^2\right\}}
\end{equation}
where $\{\tsigma_{i,n}^2,~i=0,1,\ldots,n\}$ are (possibly) mismatched noise variances
assumed by the decoder, and $\by_i$ is the $i$--th segment of $\by$, that corresponds
to the channel input segment $\bx_i[m]$.
Of course, if $\tsigma_{i,n}^2=\sigma_{i,n}^2$ for all
$i$ (or more generally, $\tsigma_{i,n}^2\propto\sigma_{i,n}^2$) and
$\beta\to\infty$, we obtain the deterministic (matched) ML decoder. Similarly
as with $\{P_{i,n}\}$, 
it will be assumed that $\tsigma_{i,n}^2=\tilde{S}_Z(e^{j2\pi i/n})$, $i=0,1,\ldots,n-1$, for some function
$\tilde{S}_Z(e^{j\omega})$, $\omega\in[0,2\pi)$, and so, together with
(\ref{psampled}) and the eigenvalue distribution theorem, we have
\begin{equation}
\lim_{n\to\infty}\frac{1}{n}\sum_{i=0}^{n-1}G(P_{i,n},\sigma_{i,n}^2,\tsigma_{i,n}^2)=
\frac{1}{2\pi}\int_{0}^{2\pi}G[S_X(e^{j\omega}),S_Z(e^{j\omega}),\tilde{S}_Z(e^{j\omega})]\mbox{d}\omega
\end{equation}
for any continuous function $G$.

Finally, the probability of error and the TRC exponent are defined as in
Subsection II.B.1,
except that now $n$ is replaced by $N$ and the GLD of eq.\ (\ref{posterior1}) is replaced by
the one of eq.\ (\ref{posterior2}). Our goal would be to derive
the TRC exponent under the asymptotic regime that we have defined. The process
of taking the limit
of $N=n\ell\to\infty$ will be carried out in two steps: we first take
the limit
$\ell\to\infty$ for fixed $n$, and then take limit $n\to\infty$. In other
words, the asymptotic regime corresponds to $\ell\gg n$.

\subsection*{C.\ Background}



As mentioned in Subsection II.B.2, eq.\ (\ref{evdt}) is motivated by the
eigenvalue distribution theorem.
Let $\{Z_i\}$ be a discrete--time, zero--mean stationary process
with an absolutely summable autocorrelation sequence $\{r_Z(k),~k=0,\pm 1,\pm 2,\ldots\}$
(i.e., $\sum_k |r_Z(k)| < \infty$), and
power spectral density
\begin{equation}
S_Z(e^{j\omega})=\calF\{r_Z(k)\}=
\sum_{k=-\infty}^{\infty}r_Z(k)e^{-j\omega k},
\end{equation}
which is assumed strictly positive and bounded for all $\omega\in[-\pi,\pi)$.
Let $R_Z^n$ be the corresponding $n\times n$ autocorrelation matrix, namely,
the matrix whose
$(k,l)$--th element is $r_Z(k-l)$,
$k,l\in\{1,\ldots,n\}$.
Let $\Lambda_n$ be the matrix whose columns are orthonormal eigenvectors
of $R_Z^n$. Then by applying the linear transformation,
$\tilde{\bZ}=\Lambda_n^{-1}\bZ$, one diagonalizes the covariance matrix and the
variances, $\{\sigma_{i,n}^2\}$, of
the various components of $\tilde{\bZ}$ are equal to the respective eigenvalues of $R_Z^n$.
If $\{r_Z(k)\}$ is absolutely summable, then
according to the eigenvalue distribution theorem, eq.\ (\ref{evdt}) holds true
\cite[Theorem 4.2]{Gray06} for every continuous function $G:[a,b]\to\reals$,
with $a=\mbox{ess}\inf S_Z(e^{j\omega})$ and
$b=\mbox{ess}\sup S_Z(e^{j\omega})$.

Applying this setup to colored Gaussian intersymbol interference (ISI) channel model, let
\begin{equation}
Y_t=\sum_{i=0}^\infty h_iX_{t-i}+W_t,
\end{equation}
where $\{W_t\}$ is a zero--mean, stationary Gaussian process with
a given spectrum $S_W(e^{j\omega})$ and 
the linear ISI system, $H(z)=\sum_{i=0}^\infty h_iz^{-i}$, has an
inverse $G(z)=1/H(z)=\sum_{i=0}^\infty g_iz^{-i}$ such that
$Z_t=\sum_{i=0}^\infty g_iW_{t-i}$ has a power spectral density,
\begin{equation}
S_Z(e^{j\omega})=S_W(e^{j\omega})|G(e^{j\omega})|^2=\frac{S_W(e^{j\omega})}{|H(e^{j\omega})|^2} 
\end{equation}
whose inverse Fourier
transform, $\{r_Z(k)\}$, is absolutely summable. Then, considering the equivalent channel
model (neglecting edge effects),
\begin{equation}
\tilde{Y}_t=X_t+Z_t,
\end{equation}
we can now apply the linear transformation $\Lambda_n$ that diagonalizes the
covariance matrix $R_Z^n$ of an $n$-block of $\bZ$, 
and then the resulting variances, $\sigma_{i,n}^2$, of the transformed noise
vector, are equal to the respective
eigenvalues of $R_Z^n$, which in turn satisfy the eigenvalue distribution
theorem in the large $n$ limit.

Operatively, the communication system works as follows: 
given a codeword $\bx=(\bx_0,\bx_1,\ldots,\bx_{n-1})$,
generated as described in Subsection II.B.2,
we transmit $\ell$ sub--blocks, each of length $n$, where in the $k$--th sub-block
($k=0,1,\ldots,\ell-1$),
the transmitted vector is $\sum_{i=0}^{n-1}x_{i\ell+k}\psi_i$, $\psi_i$ being the
$i$--th eigenvector of $R_Z^n$. At the receiver, we first apply the linear
system $G(z)$ to $\{Y_t\}$ in order to obtain $\{\tilde{Y}_t\}$. Then, every $n$--block
of $\{\tilde{Y}_t\}$ undergoes a bank of correlators with all eigenvectors,
$\{\psi_i\}$, in order to
retrieve noisy versions of the
coefficients $\{x_{in+k}\}_{i=0}^{n-1}$, 
whose noise components are uncorrelated
with each other, and their variances are $\sigma_{i,n}^2$, $i=0,1,\ldots,n-1$.

Note that, in general, $\{\psi_i\}$ and $\{x_{i\ell+k}\}$ might be complex--valued. 
For the sake of simplicity, and without loss of generality, however, we will assume 
that they are real. This is justified by the following simple consideration.
Suppose that $\psi_i$ is a complex eigenvector of $R_Z^n$. Since both $R_Z^n$ and the
corresponding eigenvalue, $\sigma_{i,n}^2$, are real, then the complex
conjugate, $\psi_i^*$ is also an eigenvector associated with $\sigma_{i,n}^2$, and therefore, so are
$\bu_i=\mbox{Re}\{\psi_i\}$ and 
$\bv_i=\mbox{Im}\{\psi_i\}$, which are vectors in $\reals^n$. Now, if $\bu_i$
and $\bv_i$ are
not orthogonal to each other, then one can apply a simple
transformation (e.g., the Gram--Schmidt projection) to represent the two--dimensional
eigensubspace, spanned by $\bu_i$ and 
$\bv_i$, by two orthonormal basis vectors, which are clearly eigenvectors
pertaining to $\sigma_{i,n}^2$ as well. Of course, eigenvectors of
other eigenvalues are orthogonal to each other. 
For example, in the
circulant approximation of $R_Z^n$ for large $n$ \cite[Chapters 3, 4]{Gray06}, 
the matrix $\Lambda_n$
pertains to the discrete Fourier transform (DFT), whose complex exponential 
eigenvectors can be separated into real--valued sine and cosine 
vectors for the real
and imaginary parts, which are all orthogonal to each other.

\section*{III.\ Main Results}

\subsection*{A.\ The AWGN Channel}

We begin from the AWGN channel model described in Subsection II.B.1. 
We first define a few quantities. For a given $\sigma_Y^2 > 0$ and
$\rho\in[-1,1]$, define
\begin{equation}
g(\sigma_Y^2,\rho)=\frac{\beta\sqrt{P}\sigma_Y\rho}{\sigma^2}
\end{equation}
and
\begin{eqnarray}
\alpha(R,\sigma_Y^2)&=&\sup_{|\rho|\le\sqrt{1-e^{-2R}}}
\left[g(\sigma_Y^2,\rho)+
\frac{1}{2}\ln(1-\rho^2)\right]+R\nonumber\\
&=&w\left(\frac{\beta\sqrt{P}\sigma_Y}{\sigma^2},R\right)+R,
\end{eqnarray}
where
\begin{eqnarray}
w(u,R)&\dfn&\sup_{|\rho|\le\sqrt{1-e^{-2R}}}
\left[\rho\cdot u+\frac{1}{2}\ln(1-\rho^2)\right]\nonumber\\
&=&\left\{\begin{array}{ll}
u\cdot\sqrt{1-e^{-2R}}-R & R \le 
\frac{1}{2}\ln\left(\frac{1+\sqrt{1+4u^2}}{2}\right)\\
\frac{2u^2}{1+\sqrt{1+4u^2}}-\frac{1}{2}\ln\left(\frac{1+\sqrt{1+4u^2}}{2}\right)
& R \ge \frac{1}{2}\ln\left(\frac{1+\sqrt{1+4u^2}}{2}\right)\end{array}\right.
\end{eqnarray}
Next, for a given $\rho_{XX^\prime}\in[-1,1]$, define
\begin{eqnarray}
\label{Gammadef}
\Gamma(\rho_{XX^\prime})&=&\frac{1}{2}\ln(2\pi\sigma^2)+\inf\left\{
\frac{1}{2\sigma^2}(\sigma_Y^2-2\sqrt{P}\sigma_Y\rho_{XY}+P)-
\frac{1}{2}\ln(2\pi e \sigma_{Y|XX^\prime}^2)+\right.\nonumber\\
& &\left.\left[\max\{g(\sigma_Y^2,\rho_{XY}),\alpha(R,\sigma_Y^2)\}-
g(\sigma_Y^2,\rho_{X^\prime
Y})\right]_+\right\},
\end{eqnarray}
where
\begin{equation}
\sigma_{Y|XX^\prime}^2\dfn\sigma_Y^2+
\min_{s,t}\{(s^2+2st\rho_{XX^\prime}+t^2)P-2\sqrt{P}\sigma_Y(s\rho_{XY}+
t\rho_{X^\prime Y})\},
\end{equation}
and where the infimum in (\ref{Gammadef}) 
is over all $\{\sigma_Y^2,\rho_{XY},\rho_{X^\prime Y}\}$
such that the matrix
\begin{equation}
\label{covmat}
\left(\begin{array}{ccc}
P & \rho_{XX^\prime}P & \rho_{XY}\sqrt{P}\sigma_Y\\
\rho_{XX^\prime}P & P & \rho_{X^\prime Y}\sqrt{P}\sigma_Y\\
\rho_{XY}\sqrt{P}\sigma_Y &  \rho_{X^\prime Y}\sqrt{P}\sigma_Y &
\sigma_Y^2\end{array}\right)
\end{equation}
is positive semi--definite.

Our first result is the following.
\begin{theorem}
\label{awgn-exponent}
For the AWGN channel model defined in Subsection II.B.1,
\begin{equation}
E_{\mbox{\tiny
trc}}(R)=\inf\left\{\Gamma(\rho_{XX^\prime})+
\frac{1}{2}\ln\frac{1}{1-\rho_{XX^\prime}^2}\right\}-R,
\end{equation}
where the infimum is subject to the constraint $|\rho_{XX^\prime}|\le\sqrt{1-e^{-4R}}$.
\end{theorem}

The proof appears in Section V.A. The remaining part of this
subsection is devoted to a discussion about Theorem 1.

The insight behind this expression of the TRC exponent is as follows. Let us
imagine auxiliary random variables $X$, $X^\prime$ 
and $\bY$, that represent the transmitted codeword
$\bx[m]$, an incorrect codeword $\bx[m^\prime]$ ($m^\prime\ne m$), and the channel output
vector, $\by$, respectively. The term
$[\max\{g(\sigma_Y^2,\rho_{XY}),\alpha(R,\sigma_Y^2)\}-g(\sigma_Y^2,\rho_{X^\prime
Y})]_+$, that appears in the definition of $\Gamma(\rho_{XX^\prime})$,
represents the exponential rate of the probability that the GLD would select
$\bx[m^\prime]$ as the decoded message given that
the empirical correlation coefficient between $\bx[m]$ and $\bx[m^\prime]$ is
a given number, denoted $\rho_{XX^\prime}$. The term $\alpha(R,\sigma_Y^2)$ is interpreted
as the typical exponent of the collective contribution of all incorrect codewords 
at the denominator of (\ref{posterior1}). As explained also in \cite{trc}, 
the probability that $\bx[m^\prime]$ would be the decoded message given that
$\bx[m]$ was transmitted, is of the exponential order of
$\exp\{-n\Gamma(\rho_{XX^\prime})\}$. Therefore, the overall error
probability, 
\begin{eqnarray}
\label{intuition}
P_{\mbox{\tiny
e}}(\calC_n)&=&\frac{1}{M}
\sum_m\sum_{m^\prime\ne m}\mbox{Pr}\left\{\bx[m^\prime]~\mbox{decoded}
\bigg|\bx[m]~\mbox{transmitted}\right\}\nonumber\\
&\exe&\sum_{\rho_{XX^\prime}}M(\rho_{XX^\prime})\exp\{-n[\Gamma(\rho_{XX^\prime})+R]\},
\end{eqnarray}
is of the exponential order of
$\max_{\rho_{XX^\prime}}M(\rho_{XX^\prime})\exp\{-n[\Gamma(\rho_{XX^\prime})+R]\}$,
where $M(\rho_{XX^\prime})$ is the typical number of codeword pairs
$(\bx[m],\bx[m^\prime])$ whose empirical correlation coefficient is about
$\rho_{XX^\prime}$ and the sum is over a fine grid in $(-1,1)$.
Since there are about $e^{2nR}$ codeword pairs in $\calC_n$ and since the
probability of event $\{\bx^T[m]\cdot\bx[m^\prime]/(nP)\approx \rho_{XX^\prime}\}$ is about
$\exp\{-\frac{n}{2}\ln[1/(1-\rho_{XX^\prime}^2)]$, the typical value of the 
number $M(\rho_{XX^\prime})$
is of the exponential order of
$\exp\{n[2R-\frac{1}{2}\ln[1/(1-\rho_{XX^\prime}^2)]\}$, whenever
$2R-\frac{1}{2}\ln[1/(1-\rho_{XX^\prime}^2)> 0$, and is zero otherwise.

An alternative representation of $\Gamma(\rho_{XX^\prime})$ is the following.
Let $X$, $X^\prime$ and $Y$ be zero--mean random variables, defined as
follows. The variables $X$ and $X^\prime$ both have variance $P$ and covariance
$\bE(XX^\prime)=\rho_{XX^\prime}P$. Given $X$ and $X^\prime$, let $Y$ be
defined as $Y=a X+b X^\prime+V$, where $V$ 
is zero--mean wit variance $\sigma_V^2$, and is
uncorrelated to both $X$ and $X^\prime$. Under this representation, we can
transform the optimization variables of $\Gamma(\rho_{XX^\prime})$, from
$\{\sigma_Y^2,\rho_{XY},\rho_{X^\prime Y}\}$ to $(a,b,\sigma_V^2)$, and then
$\Gamma(\rho_{XX^\prime})$ becomes
\begin{eqnarray}
\Gamma(\rho_{XX^\prime})&=&\frac{1}{2}\ln(2\pi\sigma^2)+\inf_{a,b,\sigma_V^2}\left(
\frac{1}{2\sigma^2}([(a-1)^2+2\rho_{XX^\prime}(a-1)b+b^2]P+\sigma_V^2)-
\frac{1}{2}\ln(2\pi e\sigma_V^2)+\right.\nonumber\\
& &\left.\left[\max\left\{\frac{\beta}{\sigma^2}(a+\rho_{XX^\prime}
b)P,\alpha(R,[a^2+2\rho_{XX^\prime}ab+b^2]P+\sigma_V^2)\right\}-
\frac{\beta}{\sigma^2}(\rho_{XX^\prime} a +b)P\nonumber
\right]_+\right),
\end{eqnarray}
where the infimum is over $(a,b,\sigma_V^2)\in\reals^2\times\reals^+$.

We observe that by eliminating the term
$\alpha(R,[a^2+2\rho_{XX^\prime}ab+b^2]P+\sigma_V^2)$, the expression of
$\Gamma(\rho_{XX^\prime})$ simplifies dramatically, and hence also the
expression of the TRC exponent. As we show next, in this case, all the
minimizations can be carried out in closed form. 
This elimination of the term $\alpha(R,\cdot)$
yields a lower bound to the TRC exponent, which corresponds to a union bound
of the pairwise error events, $\{m\to m^\prime\}$, and which is tight at a
certain range of low rates. Let us define then
\begin{eqnarray}
\label{GammaLoptimization}
\Gamma_L(\rho_{XX^\prime})&\dfn&\frac{1}{2}\ln(2\pi\sigma^2)+\inf_{a,b,\sigma_V^2}\left[
\frac{1}{2\sigma^2}([(a-1)^2+2\rho_{XX^\prime}(a-1)b+b^2]P+\sigma_V^2)-\right.\nonumber\\
& &\left.\frac{1}{2}\ln(2\pi e\sigma_V^2)+
\frac{\beta P}{\sigma^2}[a+\rho_{XX^\prime}
b-\rho_{XX^\prime} a -b)
]_+\right]\nonumber\\
&=&\frac{1}{2}\ln(2\pi\sigma^2)+\inf_{a,b,\sigma_V^2}\left[
\frac{1}{2\sigma^2}([(a-1)^2+2\rho_{XX^\prime}(a-1)b+b^2]P+\sigma_V^2)-\right.\nonumber\\
& &\left.\frac{1}{2}\ln(2\pi e\sigma_V^2)+
\frac{\beta P}{\sigma^2}(1-\rho_{XX^\prime})[a-b]_+
\right]\nonumber\\
&=&\frac{1}{2}\ln(2\pi\sigma^2)+\inf_{a,b,\sigma_V^2}\sup_{0\le\lambda\le\beta}\left[
\frac{1}{2\sigma^2}([(a-1)^2+2\rho_{XX^\prime}(a-1)b+b^2]P+\sigma_V^2)-\right.\nonumber\\
& &\left.\frac{1}{2}\ln(2\pi e\sigma_V^2)+
\frac{\lambda P}{\sigma^2}\cdot(1-\rho_{XX^\prime})(a-b)
\right]\nonumber\\
&\eqa&\frac{1}{2}\ln(2\pi\sigma^2)+\sup_{0\le\lambda\le\beta}\inf_{a,b,\sigma_V^2}\left[
\frac{1}{2\sigma^2}([(a-1)^2+2\rho_{XX^\prime}(a-1)b+b^2]P+\sigma_V^2)-\right.\nonumber\\
& &\left.\frac{1}{2}\ln(2\pi e\sigma_V^2)+
\frac{\lambda P}{\sigma^2}(1-\rho_{XX^\prime})(a-b)
\right]\nonumber\\
&\eqb&\sup_{0\le\lambda\le\beta}\inf_{a,b}\left[
\frac{P}{2\sigma^2}\cdot[(a-1)^2+2\rho_{XX^\prime}(a-1)b+b^2]+\right.\nonumber\\
& &\left.\frac{\lambda P}{\sigma^2}(1-\rho_{XX^\prime})(a-b)
\right]\nonumber\\
&=&\sup_{0\le\lambda\le\beta}
\frac{P}{\sigma^2}\cdot\lambda(1-\lambda)(1-\rho_{XX^\prime})\nonumber\\
&=&\frac{P}{\sigma^2}\cdot\hat{\beta}(1-\hat{\beta})\cdot(1-\rho_{XX^\prime}),
\end{eqnarray}
where $\hat{\beta}=\min\{\beta,\frac{1}{2}\}$. Here, the passage (a) is allowed
by the minimax theorem as the objective is convex in $(a,b,\sigma_V^2)$ and
affine (and hence concave) in $\lambda$ and passage (b) is by the simple fact
that the optimal $\sigma_V^2$ turns out to be equal to $\sigma^2$.
Thus, for $\beta\ge
\frac{1}{2}$,
\begin{equation}
\Gamma_L(\rho_{XX^\prime})=\frac{P}{4\sigma^2}\cdot(1-\rho_{XX^\prime})\dfn
\frac{\mbox{snr}}{4}\cdot(1-\rho_{XX^\prime}).
\end{equation}
Finally, 
\begin{eqnarray}
\label{awgnlowerbound}
E_{\mbox{\tiny
trc}}(R)&\ge& E_{\mbox{\tiny trc}}^-(R)\nonumber\\
&\dfn&\inf_{|\rho_{XX^\prime}|\le\sqrt{1-e^{-4R}}}
\left[\frac{\mbox{snr}}{4}\cdot(1-\rho_{XX^\prime})
-\frac{1}{2}\ln(1-\rho_{XX^\prime}^2)-R\right]\nonumber\\
&=&\frac{\mbox{snr}}{4}-w\left(\frac{\mbox{snr}}{4},2R\right)-R\nonumber\\
&=&\left\{\begin{array}{ll}
\frac{\mbox{snr}}{4}(1-\sqrt{1-e^{-4R}})+R & R \le R_*\\
\frac{\mbox{snr}}{4}-\frac{\mbox{snr}^2/8}{1+\sqrt{1+\mbox{snr}^2/4}}+\frac{1}{2}
\ln\left(\frac{1+\sqrt{1+\mbox{snr}^2/4}}{2}\right)-R & R \ge R_*\end{array}\right.
\end{eqnarray}
where
\begin{equation}
\label{Rstar}
R_*=\frac{1}{4}\ln\left(\frac{1+\sqrt{1+\mbox{snr}^2/4}}{2}\right).
\end{equation}
The lower bound, $E_{\mbox{\tiny trc}}^-(R)$,
has a non--affine convex part, starting with slope
$-\infty$ at rate $R=0$, and ending with
slope $-1$ at rate $R_*$, and so, it is tangential to the straight--line part 
that starts at rate $R_*$. The point $R=R_*$ exhibits a {\it glassy 
phase transition} \cite[Chap.\ 6]{spit}, \cite[Chapters 5,6]{MM09} in the behavior of
$E_{\mbox{\tiny trc}}^-(R)$: for $R
\le R_*$, the error probability (see second line of eq.\ (\ref{intuition}))
is dominated by a sub--exponential number of
incorrect codewords at Euclidean distance $d=\sqrt{2nP(1-\sqrt{1-e^{-4R}})}$ from the
transmitted codeword, 
whereas for $R>R_*$, there are exponentially
$\exp\{n[2R-\frac{1}{2}\ln\frac{1}{1-\rho_*^2}]\}=\exp\{2n(R-R_*)\}$ dominant codewords at
Euclidean distance $d=\sqrt{2nP(1-\rho^*)}$, where
$\rho^*=\mbox{snr}/[2(1+\sqrt{1+\mbox{snr}^2/4})]$.
The straight--line part
of eq.\ (\ref{awgnlowerbound}) can 
readily be recognized as $R_0-R$, which is exactly the straight--line part of
the random coding error exponent below the critical rate. Thus, the TRC
exponent exceeds the random coding exponent at least in the range
$R\in[0,R_*]$. Obviously, for $R\ge R_{\mbox{\tiny crit}}\ge R_*$, the exact TRC exponent
must coincide with the random coding exponent, 
as the TRC exponent is sandwiched between the
random coding exponent and the sphere--packing exponent, which in turn
coincide for $R\ge R_{\mbox{\tiny crit}}$. Thus, the interesting range to
explore the behavior of the TRC exponent is the range of low rates. 

At the low--rate extreme, it is readily observed that the lower bound
(\ref{awgnlowerbound}) yields $E_{\mbox{\tiny
trc}}(0)\ge\mbox{snr}/4$, which must be, in fact, an equality, since it coincides with
the minimum--distance upper bound on the best zero--rate achievable error
exponent. In
this context, a natural question that arises is the following: what is
the range of low rates 
where the above derived lower bound to 
the TRC exponent is tight, i.e., $E_{\mbox{\tiny trc}}^-(R)=E_{\mbox{\tiny
trc}}(R)$? In other words, at what range of rates, the union
bound of the pairwise error probabilities is of the same exponential order as the exact
TRC exponent? 
We will propose an answer to this question in more generality in
the sequel, where we consider the colored Gaussian noise model.

\subsection*{B.\ Parallel Additive Gaussian Channels and the Colored Gaussian Channel}

We next move on to handle the model of independent parallel Gaussian channels, which
will then be used pass to the colored Gaussian channel, as described above.
Similarly as in the case of the AWGN channel, considered 
in Subsection III.A, here too, the elimination of the term $\alpha(R,\cdot)$
contributes dramatically to the simplification of the ultimate TRC exponent
formula (lower bound). Moreover, without it, 
the resulting expression would be associated with a very complicated
calculus of variations. Also, as before, this simplification comes at no loss
of tightness at some range of low rates, as will be shown in the sequel. 

Consider the setup defined in Subsection II.B.2, and to avoid cumbersome
notation, we henceforth omit the subscript $n$ of $P_{i,n}$, $\sigma_{i,n}^2$, and
$\tsigma_{i,n}^2$, and denote them instead by $P_i$, $\sigma_i^2$, and
$\tsigma_i^2$, respectively. We will also use the notation
$\mu_i=\sigma_i^2/\tsigma_i^2$,
and $\mbox{snr}_i=P_i/\sigma_i^2$. 
For a given $\lambda \ge 0$ and $\theta \ge 1$, let us define
\begin{equation}
A(\mbox{snr}_i,\mu_i,\lambda,\theta)\dfn\inf_{|\rho|< 1}\left\{
\lambda\mu_i(1-\lambda\mu_i)\mbox{snr}_i(1-\rho)+
\frac{\theta}{2}\ln\frac{1}{1-\rho^2}\right\}.
\end{equation}
More explicitly, denoting
\begin{equation}
S_i=\lambda\mu_i(1-\lambda\mu_i)\mbox{snr}_i,
\end{equation}
the minimizing $\rho$ is given by
\begin{equation}
\rho_i^*=\frac{\sqrt{\theta^2+4S_i^2}-\theta}{2S_i}=
\frac{2S_i}{\sqrt{\theta^2+4S_i^2}+\theta},
\end{equation}
and then
\begin{equation}
A(\mbox{snr}_i,\mu_i,\lambda,\theta)=
S_i\left(1-
\frac{2S_i}{\sqrt{\theta^2+4S_i^2}+\theta}\right)-
\frac{\theta}{2}\ln\left[\frac{2\theta}{\sqrt{\theta^2+4S_i^2}+\theta}\right].
\end{equation}

Our first main result in this subsection is provided by the following
theorem, whose proof appears in Section V.B.
\begin{theorem}
\label{thm2}
For the model of the parallel Gaussian channels described in Subsection
II.B.2, let $\ell\to\infty$ while $n$ is kept fixed. Then,
\begin{equation}
E_{\mbox{\tiny trc}}(R)\ge\sup_{\theta\ge
1}\sup_{0\le\lambda\le\beta}\left\{
\frac{1}{n}\sum_{i=0}^{n-1}A(\mbox{snr}_i,\mu_i,\lambda,\theta)-(2\theta-1)R\right\}.
\end{equation}
\end{theorem}

Let $S_X(\omega)$, $S_Z(\omega)$, and $\tilde{S}_Z(\omega)$ be defined as in
Subsection II.B.2, and define 
$\mbox{snr}(\omega)=S_X(\omega)/S_Z(\omega)$ and
$\mu(\omega)=S_Z(\omega)/\tilde{S}_Z(\omega)$.
The following corollary follows from Theorem \ref{thm2} using the eigenvalue distribution
theorem, by taking the limit $n\to\infty$.

\begin{corollary}
For the colored Gaussian channel described in Subsection II.B.2,
\begin{equation}
E_{\mbox{\tiny trc}}(R)\ge E_{\mbox{\tiny trc}}^-(R)\dfn\sup_{\theta\ge
1}\sup_{0\le\lambda\le\beta}\left\{
\frac{1}{2\pi}\int_{0}^{2\pi}A[\mbox{snr}(\omega),\mu(\omega),\lambda,\theta]
\mbox{d}\omega-(2\theta-1)R\right\}.
\end{equation}
\end{corollary}

Referring to Corollary 1, let us denote
\begin{equation}
B(\theta)=\sup_{0\le\lambda\le\beta}\frac{1}{2\pi}
\int_{0}^{2\pi}A[\mbox{snr}(\omega),\mu(\omega),\lambda,\theta]
\mbox{d}\omega,
\end{equation}
so that
\begin{equation}
E_{\mbox{\tiny trc}}^-(R)=\sup_{\theta\ge 1}[B(\theta)-(2\theta-1)R].
\end{equation}
A few comments are in order concerning these results.\\

\noindent
{\bf 1. The matched case.}
Note that in the matched case ($\mu_i\equiv 1$ for parallel
channels, or $\mu(\omega)\equiv 1$ for the colored channel), the
optimal value\footnote{Of course, any positive constant $c$, 
for which $\mu_i\equiv c$ (or
$\mu(\omega)\equiv c$), is also associated with matched decoding, but
the optimization over $\lambda$ would absorb such a constant.}
of $\lambda$ is $\hat{\beta}=\min\{\beta,\frac{1}{2}\}$. This simplifies the
formula in the sense that it remains to maximize (numerically) over one parameter only --
the parameter $\theta$. It also implies that for any $\beta\ge\frac{1}{2}$, the
GLD is as good as the (deterministic) ML decoder in terms of (our lower
bound to) the TRC exponent.\\

\noindent
{\bf 2. General properties of the TRC exponent function.}
The behavior of the function $E_{\mbox{\tiny trc}}^-(R)$ is similar to the that of the
AWGN case. We first observe that
$E_{\mbox{\tiny trc}}^-(0)=\sup_{\theta\ge 1}B(\theta)=\lim_{\theta\to
\infty}B(\theta)$ since $B(\theta)$ is
a monotonically non-decreasing function.
For low positive rates, $E_{\mbox{\tiny trc}}^-(R)$ is a convex curve, with 
an initial slope of $-\infty$. In this range, it can be represented
parametrically (with a slight abuse of notation) by the pair of equations,
\begin{eqnarray}
R(\theta)&=&\frac{1}{2}\cdot\frac{\mbox{d}B(\theta)}{\mbox{d}\theta}\dfn
\frac{B^\prime(\theta)}{2}\\
E_{\mbox{\tiny trc}}^-(\theta)&=&B(\theta)-(2\theta-1)R(\theta),
\end{eqnarray}
where $\theta$ exhausts the range $[1,\infty)$.
Since $B(\theta)$ is non--decreasing
and concave (because it is obtained as the minimum over affine functions of $\theta$),
then as $R$ increases,
the negative slope of $E_{\mbox{\tiny trc}}(R)$ becomes milder: it is
given by $-(2\theta_R-1)$, where $\theta_R$ is the solution $\theta$ to the
equation $B^\prime(\theta)=2R$. Since $B^\prime(\theta)$ is a decreasing
function (due to the concavity of $B(\theta)$), then so is $\theta_R$. 
The curvy part of $E_{\mbox{\tiny trc}}^-(R)$ ends
at the point where $\theta_R=1$. This happens at rate 
$R_*=B^\prime(1)/2$. For $R\ge R_*$,
$E_{\mbox{\tiny trc}}^-(R)=B(1)-R$,
where $B(1)$ is given by
\begin{equation}
B(1)=\sup_{0\le\lambda\le\beta}\frac{1}{2\pi}\int_0^{2\pi}A[\mbox{snr}(\omega),\mu(\omega),
\lambda,1]\mbox{d}\omega.
\end{equation}
It is shown in Appendix B that this expression coincides also with $R_0$, the
zero--rate random coding exponent, and so, in the range of rates between
$R_*$ and $R_{\mbox{\tiny crit}}$, our lower bound to the TRC exponent
coincides with the classical random coding exponent.

To calculate $R_*$, we have
\begin{eqnarray}
R_*&=&\frac{1}{2}\cdot\frac{\partial}{\partial\theta}
\left[\frac{1}{2\pi}\int_{0}^{2\pi}A[\mbox{snr}(\omega),\mu(\omega),
\lambda,\theta]\mbox{d}\omega
\right]_{\theta=1}\nonumber\\
&=&\frac{1}{4\pi}\int_{0}^{2\pi}\left(\frac{\partial}{\partial\theta}
A[\mbox{snr}(\omega),\mu(\omega),\lambda,\theta]\right)_{\theta=1}\mbox{d}\omega\nonumber\\
&=&\frac{1}{4\pi}\int_{0}^{2\pi}\frac{\partial}{\partial\theta}
\left[\lambda\mu(\omega)[1-\lambda\mu(\omega)]\mbox{snr}(\omega)
(1-\rho_{\theta})+\frac{\theta}{2}
\ln\frac{1}{1-\rho_{\theta}^2}\right]_{\theta=1}\mbox{d}\omega\nonumber\\
&\eqa&\frac{1}{4\pi}\int_{0}^{2\pi}\frac{1}{2}\ln\frac{1}{1-
\rho_{1}^2}\mbox{d}\omega\nonumber\\
&=&\frac{1}{8\pi}\int_{0}^{2\pi}
\ln\left[\frac{1+\sqrt{1+4\lambda^2\mu^2(\omega)[1-
\lambda\mu(\omega)]^2\mbox{snr}^2(\omega)}}{2}\right]\mbox{d}\omega,
\end{eqnarray}
where $\lambda$ is the achiever of the TRC exponent and $\rho_\theta$ is the
optimal $\rho$ for a given $\theta$. Equality (a) is obtained by observing
that upon differentiating the integrand, the internal derivative 
$\mbox{d}\rho_\theta/\mbox{d}\theta$ is multiplied by an expression that
vanishes due to the fact that $\rho_\theta$ (which is also a function of
$\omega$) is optimal.
In the matched case
($\mu(\omega)\equiv 1$, which is ML
decoding), where the optimal value of $\lambda$ is $\frac{1}{2}$, this becomes
\begin{equation}
R_*=\frac{1}{8\pi}\int_{0}^{2\pi}
\ln\left[\frac{1+\sqrt{1+\mbox{snr}^2(\omega)/4}}
{2}\right]\mbox{d}\omega,
\end{equation}
thus recovering the expression (\ref{Rstar}) of $R_*$ of the AWGN channel as a special case.
Obviously, $R_{\mbox{\tiny crit}}$ must be larger than $R_*$, otherwise, the
TRC exponent would exceed the sphere--packing exponent along the range
$[R_{\mbox{\tiny crit}},R_*]$, which is a clear contradiction.\\

\noindent
{\bf 3. Tightness at low rates.} Recall that in our analysis, both here and in
the AWGN channel case, we have ignored the term
$\alpha(R,\cdot)$ that designates the contribution of all the incorrect
codewords in the posterior of the GLD (see eqs.\ (\ref{posterior1}) and
(\ref{posterior2})). It turns out that at least for deterministic mismatched
decoding at low rates, this simplification comes
at no cost in the exponential tightness. More precisely, there is an interval of
low rates $[0,R_{\mbox{\tiny t}}]$, where 
$E_{\mbox{\tiny trc}}^-(R)=E_{\mbox{\tiny trc}}(R)$. In Appendix C, we derive
a non--trivial lower bound to $R_{\mbox{\tiny t}}$.\\

\noindent
{\bf 4. The zero--rate TRC exponent.}
For $R=0$, we have
\begin{eqnarray}
E_{\mbox{\tiny trc}}^-(0)
&=&\sup_{\theta\ge 1}\sup_{0\le\lambda\le
\beta}\frac{1}{2\pi}\int_0^{2\pi}A[\mbox{snr}(\omega),\mu(\omega),\lambda,\theta]
\mbox{d}\omega\nonumber\\
&=&\sup_{0\le\lambda\le
\beta}\lim_{\theta\to\infty}\frac{1}{2\pi}\int_0^{2\pi}
A[\mbox{snr}(\omega),\mu(\omega),\lambda,\theta]
\mbox{d}\omega\nonumber\\
&=&\sup_{0\le\lambda\le\beta}\frac{1}{2\pi}\int_{0}^{2\pi}
\lambda\mu(\omega)
[1-\lambda\mu(\omega)]\mbox{snr}(\omega)\mbox{d}\omega\nonumber\\
&=&\sup_{0\le\lambda\le\beta}\left\{\lambda\cdot\frac{1}{2\pi}\int_{0}^{2\pi} 
\mu(\omega)\cdot\mbox{snr}(\omega)\mbox{d}\omega-
\lambda^2\cdot\frac{1}{2\pi}\int_{0}^{2\pi}\mu^2(\omega)\cdot
\mbox{snr}(\omega)\mbox{d}\omega\right\}.
\end{eqnarray}
Now, if 
\begin{equation}
\beta < \frac{\int_0^{2\pi}\mu(\omega)\cdot\mbox{snr}(\omega)\mbox{d}\omega}{
2\int_0^{2\pi}\mu^2(\omega)\cdot\mbox{snr}(\omega)\mbox{d}\omega}
\end{equation}
then 
\begin{equation}
E_{\mbox{\tiny trc}}^-(0)=\beta\cdot\frac{1}{2\pi}\int_{0}^{2\pi} 
\mu(\omega)\cdot\mbox{snr}(\omega)\mbox{d}\omega-
\beta^2\cdot\frac{1}{2\pi}\int_{0}^{2\pi}\mu^2(\omega)\cdot
\mbox{snr}(\omega)\mbox{d}\omega.
\end{equation}
Otherwise,
the optimal $\lambda$ is given by
\begin{equation}
\lambda^*=\frac{\int_0^{2\pi}\mu(\omega)\cdot\mbox{snr}(\omega)\mbox{d}\omega}{
2\int_0^{2\pi}\mu^2(\omega)\cdot\mbox{snr}(\omega)\mbox{d}\omega},
\end{equation}
and then
\begin{equation}
E_{\mbox{\tiny
trc}}^-(0)=\frac{1}{8\pi}\cdot\frac{\left[\int_0^{2\pi}
\mu(\omega)\cdot\mbox{snr}(\omega)\mbox{d}\omega\right]^2}
{\int_0^{2\pi}\mu^2(\omega)\cdot\mbox{snr}(\omega)\mbox{d}\omega}.
\end{equation}
Observe that by the Cauchy--Shwartz inequality
\begin{eqnarray}
& &\frac{1}{8\pi}\cdot\frac{\left[\int_{0}^{2\pi}\mu(\omega)\cdot
\mbox{snr}(\omega)\mbox{d}\omega\right]^2}{\int_{0}^{2\pi}
\mu^2(\omega)\cdot\mbox{snr}(\omega)\mbox{d}\omega}\nonumber\\
&=&\frac{1}{8\pi}\cdot\frac{\left[\int_{0}^{2\pi}
\sqrt{\mu^2(\omega)\cdot\mbox{snr}(\omega)}\cdot
\sqrt{\mbox{snr}(\omega)}\mbox{d}\omega\right]^2}{\int_{0}^{2\pi}
\mu^2(\omega)\cdot\mbox{snr}(\omega)\mbox{d}\omega}\nonumber\\
&\le&\frac{1}{8\pi}\cdot\frac{\int_{0}^{2\pi}\mu^2(\omega)\cdot
\mbox{snr}(\omega)\mbox{d}\omega\cdot
\int_{0}^{2\pi}
\mbox{snr}(\omega)\mbox{d}\omega}{\int_{0}^{2\pi}
\mu^2(\omega)\cdot\mbox{snr}(\omega)\mbox{d}\omega}\nonumber\\
&=&\frac{1}{8\pi}\cdot\int_{0}^{2\pi}
\mbox{snr}(\omega)\mbox{d}\omega,
\end{eqnarray}
with equality iff $\mu(\omega)=\mbox{const.}$ for almost every $\omega$,
which is the matched case.\\

\noindent
{\bf 5. The continuous--time colored Gaussian channel.}
A very similar analysis applies to the continuous--time colored Gaussian channel. Here, we
begin from $\ell$ non--overlapping 
frames, each of length $T_0$ seconds. The resulting expression
of $\log P_{\mbox{\tiny e}}(C_n)$,
using the very same ideas, would be of the general form
$\ell\sup\sum_iG[\lambda_i(T_0)]$, where
$\lambda_i(T_0)$ are the eigenvalues pertaining
to the noise autocorrelation function (see \cite[Chap.\ 8]{Gallager68}). But
$$\ell\cdot\sup\sum_iG[\lambda_i(T_0)]=
\ell T_0\cdot\sup\frac{1}{T_0}\sum_iG[\lambda_i(T_0)],$$
which for large $T_0$, behaves like
$$\ell T_0\cdot\sup\frac{1}{2\pi}\int_{-\infty}^{+\infty}G[S_Z(\omega)]\mbox{d}\omega
=T\cdot\sup\frac{1}{2\pi}\int_{-\infty}^{+\infty}G[S_Z(\omega)]\mbox{d}\omega,$$
where $T=\ell T_0$ is the overall duration of all $\ell$ frames of length $T_0$. Thus,
in the continuous--time case, we obtain exactly the same error exponent formula,
except for 
two differences: (i) $\omega$ is now analog frequency (in units of radians per
second, rather
than just radians), and the range of all frequency--domain integrals is from
$-\infty$ to $+\infty$, rather than from $0$ to $2\pi$.
(ii) The exponent is in terms of the duration $T$,
rather than the integer $N$, namely, $P_{\mbox{\tiny e}}(\calC_n)\exe e^{-ET}$ rather
than $P_{\mbox{\tiny e}}(\calC_n)\exe e^{-En}$.

\section*{IV.\ Water--Pouring Optimization of the Input Power Spectrum}

A natural question that always arises in the context of parallel Gaussian
channels and the colored Gaussian channel is the question of the optimal input
spectrum. This section is devoted to address this question in the context of
the TRC exponent. 

The function $A(\mbox{snr},\mu,\lambda,\theta)$
is concave in $\mbox{snr}$ (as it is the minimization of affine functions of
$\mbox{snr}$ over the parameter $\rho$), and considering first, the parallel Gaussian channels, 
we would like to maximize $\sum_{i=0}^{n-1}A(\mbox{snr}_i,\mu_i,\lambda,\theta)$
subject to the constraints, $\sum_{i=0}^{n-1}\sigma_i^2\mbox{snr}_i\le nP$ and
$\mbox{snr}_i\ge 0$ for all $i$. This amounts to the Lagrangian,
\begin{equation}
\sum_{i=0}^{n-1}A(\mbox{snr}_i,\mu_i,\lambda,\theta)+
\xi\left(nP-\sum_{i=0}^{n-1}\sigma_i^2\mbox{snr}_i\right)+\sum_{i=0}^{n-1}\nu_i\mbox{snr}_i.
\end{equation}
Denoting by $\dot{A}(\mbox{snr}_i,\mu_i,\lambda,\theta)$ the partial derivative of
$A(\mbox{snr}_i,\mu_i,\lambda,\theta)$ w.r.t. $\mbox{snr}_i$,
the conditions for optimality are
\begin{equation}
\dot{A}(\mbox{snr}_i,\mu_i,\lambda,\theta)\le\xi\sigma_i^2,~~~~~~i=0,1,\ldots,n-1
\end{equation}
with equality whenever $\mbox{snr}_i > 0$. The solution is
\begin{equation}
\mbox{snr}_i=\left[\dot{A}^{-1}(\xi\sigma_i^2,\mu_i,\lambda,\theta)\right]_+,
\end{equation}
where $\dot{A}^{-1}(\cdot,\mu_i,\lambda,\theta)$ is the inverse\footnote{Since
$A$ is concave in $\mbox{snr}$, its derivative monotonically decreasing and
hence the inverse exists.} of
$\dot{A}(\cdot,\mu_i,\lambda,\theta)$ and 
$\xi\ge \dot{A}(0,\mu_i,\lambda,\theta)/\sigma_{\min}^2$ 
(with $\sigma_{\min}^2\dfn\min_{i:~\mbox{snr}_i=0}\sigma_i^2$) is
chosen such that
\begin{equation}
\sum_{i=0}^{n-1}\sigma_i^2\left[\dot{A}^{-1}(\xi\sigma_i^2,\mu_i,\lambda,\theta)\right]_+=nP.
\end{equation}
We can now easly pass to continuous--frequency integrals as before and obtain
\begin{equation}
S_X(\omega)=S_Z(\omega)\cdot\left[\dot{A}^{-1}(\xi
S_Z(\omega),\mu(\omega),\lambda,\theta)\right]_+,
\end{equation}
and finally maximize the resulting error--exponent 
expression over $\lambda$ and $\theta$. 
More specifically, using the concrete form of the function
$A$, we have:
\begin{eqnarray}
A(\mbox{snr},\mu,\lambda,\theta)&=&
\min_\rho\left\{\lambda\mu(1-\lambda\mu)\mbox{snr}(1-\rho)+\frac{\theta}{2}
\cdot\ln\frac{1}{1-\rho^2}\right\}\nonumber\\
&=&\lambda\mu(1-\lambda\mu)\mbox{snr}\cdot[1-\rho(\mbox{snr})]+
\frac{\vartheta}{2}\cdot\ln\frac{1}{1-\rho^2(\mbox{snr})},
\end{eqnarray}
which leads to
\begin{eqnarray}
\dot{A}(\mbox{snr},\mu,\lambda,\theta)&=&\lambda\mu(1-\lambda\mu)[1-\rho(\mbox{snr})]+
\left[-\lambda\mu(1-\lambda\mu)\mbox{snr}+
\frac{\theta\rho(\mbox{snr})}{1-\rho^2(\mbox{snr})}\right]
\cdot\frac{\mbox{d}\rho(\mbox{snr})}{\mbox{d}\mbox{snr}}\nonumber\\
&=&\lambda\mu(1-\lambda\mu)[1-\rho(\mbox{snr})],
\end{eqnarray}
where the second equality is due to the fact that the expression in the square
brackets, which is the derivative of $A$ w.r.t.\ $\rho$, must vanish for
$\rho(\mbox{snr})$, the optimal value of $\rho$.
Thus, the optimality condition is
\begin{equation}
\lambda\mu(1-\lambda\mu)[1-\rho(\mbox{snr}_i)]\le \xi\sigma_i^2,
\end{equation}
with equality when $\mbox{snr}_i > 0$, 
or, equivalently,
\begin{equation}
\rho(\mbox{snr}_i)=\frac{2\lambda\mu(1-\lambda\mu)\mbox{snr}_i/2}
{\theta+\sqrt{\theta^2+4\lambda^2\mu^2(1-\lambda\mu)^2\mbox{snr}_i^2}}
= \left[1-\frac{\xi\sigma_i^2}{\lambda\mu(1-\lambda\mu)}\right]_+.
\end{equation}
Upon solving for $\mbox{snr}_i$, one obtains
\begin{equation}
\mbox{snr}_i=\frac{1}{\lambda\mu_i(1-\lambda\mu_i)}\cdot
\frac{\theta[1-\xi\sigma_i^2/\{\lambda\mu_i(1-\lambda\mu_i)\}]_+}
{1-[1-\xi\sigma_i^2/\{\lambda\mu_i(1-\lambda\mu_i)\}]_+^2}.
\end{equation}
Passing to the limit of the continuous frequency domain, we obtain
\begin{eqnarray}
S_X(\omega)&=&\frac{S_Z(\omega)}{\lambda\mu(\omega)[1-\lambda\mu(\omega)]}\cdot
\frac{\theta[1-\xi S_Z(\omega)/\{\lambda\mu(\omega)[1-\lambda\mu(\omega)])\}]_+}
{1-[1-\xi
S_Z(\omega)/\{\lambda\mu(\omega)[1-\lambda\mu(\omega)]\}]_+^2}\nonumber\\
&=&\left\{\begin{array}{ll}
\frac{\theta\{\lambda\mu(\omega)[1-\lambda\mu(\omega)]-\xi S_Z(\omega)\}}
{2\xi\lambda\mu(\omega)[1-\lambda\mu(\omega)]-\xi^2S_Z(\omega)} & \xi
S_Z(\omega) \le \lambda\mu(\omega)[1-\lambda\mu(\omega)]\\
0 & \mbox{elsewhere}\end{array}\right.\nonumber\\
\end{eqnarray}
or, denoting $B=1/4\xi$,
\begin{eqnarray}
S_X(\omega)&=&\left\{\begin{array}{ll}
\frac{4\theta B\{4B\lambda\mu(\omega)[1-\lambda\mu(\omega)]-S_Z(\omega)\}}
{8B\lambda\mu(\omega)[1-\lambda\mu(\omega)]-S_Z(\omega)} & 
S_Z(\omega) \le 4B\lambda\mu(\omega)[1-\lambda\mu(\omega)]\\
0 & \mbox{elsewhere}\end{array}\right.\nonumber\\
&=&\frac{4\theta B[4B\lambda\mu(\omega)\{1-\lambda\mu(\omega)\}-S_Z(\omega)]_+}
{4B\lambda\mu(\omega)[1-\lambda\mu(\omega)]+[4B\lambda\mu(\omega)
\{1-\lambda\mu(\omega)\}-S_Z(\omega)]_+},
\end{eqnarray}
where $B$ is chosen such that
\begin{equation}
\frac{1}{2\pi}\int_{0}^{2\pi}
\frac{4B[4B\lambda\mu(\omega)\{1-\lambda\mu(\omega)\}-S_Z(\omega)]_+}
{4B\lambda\mu(\omega)[1-\lambda\mu(\omega)]+[4B\lambda\mu(\omega)
\{1-\lambda\mu(\omega)\}-S_Z(\omega)]_+}\cdot\mbox{d}\omega=\frac{P}{\theta}.
\end{equation}
Note that the optimum input spectrum depends on $R$, via the variable
$\theta$, whose optimal value depends on $R$.
When $\theta\to\infty$ ($R\to 0$), the r.h.s.\ goes to zero, and $B$ must be
chosen just slightly above $\min_\omega
S_Z(\omega)/\{\lambda\mu(\omega)[1-\lambda\mu(\omega)]\}$, in
order to comply with the power constraint. This means that $S_X(\omega)$
would tend to concentrate all the power at the frequency $\omega^*$, which
achieves this minimum.

In the matched case, where $\lambda\mu(\omega)\equiv 1/2$, we obtain
\begin{equation}
S_X(\omega)=\frac{4\theta B[B-S_Z(\omega)]_+}{B+[B-S_Z(\omega)]_+}.
\end{equation}
This optimal spectral power distribution is identical to
that of the ordinary expurgated exponent for parallel additive Gaussian
channel (see \cite[eq.\ (7.5.51), p.\ 352]{Gallager68}).
This should not be very surprising as the expurgated exponent and the TRC
exponent are closely related \cite{BF02}, \cite{trc}.

\section*{V.\ Proofs}

For the proofs, we will need some additional notation and a few preliminary
facts. 

The empirical variance of a sequence $\bx\in\reals^n$
is defined as
\begin{equation}
\hsigma_{\bx}^2=\frac{1}{n}\sum_{i=1}^nx_i^2=\frac{\|\bx\|^2}{n}.
\end{equation}
For a given $\epsilon > 0$,
the Gaussian type class of $\bx\in\reals^n$ with tolerance $\epsilon$,
will be defined as
\begin{equation}
\calT_{\epsilon}(\bx)=\left\{
\bx^\prime:~|\hsigma_{\bx'}^2-\hsigma_{\bx}^2|\le\epsilon\right\}. 
\end{equation}
The differential entropy associated with
$\hsigma_{\bx}^2$, which is the Gaussian empirical entropy of $\bx$, will be
defined as
\begin{equation}
\hat{h}_{\bx}(X)=\frac{1}{2}\ln(2\pi e\hsigma_{\bx}^2). 
\end{equation}
Similar conventions will apply to conditional empirical
conditional types and empirical conditional Gaussian differential entropies
and mutual informations.
The empirical covariance of $(\bx,\by)\in\reals^n\times\reals^n$ will be
defined as
\begin{equation}
\hC_{\bx\by}=\frac{1}{n}\sum_{i=1}^nx_iy_i=\frac{\bx^T\by}{n},
\end{equation}
The empirical correlation coefficient of
$(\bx,\by)\in\reals^n\times\reals^n$ will be defined as
\begin{equation}
\hat{\rho}_{\bx\by}=\frac{\hC_{\bx\by}}{\hsigma_{\bx}\hsigma_{\by}}.
\end{equation}
Accordingly, for a given $\epsilon > 0$, the
Gaussian conditional type class of $\by\in\reals^n$ given
$\bx\in\reals^n$,
with tolerance $\epsilon$, will be defined as
\begin{equation}
\calT_{\epsilon}(\by|\bx)=
\left\{\by^\prime:~|\hsigma_{\by^\prime}^2-\hsigma_{\by}^2|\le\epsilon,~
|\hC_{\bx\by^\prime}-\hC_{\bx\by}|\le\epsilon\right\}.
\end{equation}
The Gaussian empirical conditional entropy of $\by$ given $\bx$ will be
defined as
\begin{equation}
\hat{h}_{\bx\by}(Y|X)=\frac{1}{2}\ln[2\pi
e\hsigma_{\by}^2(1-\hat{\rho}_{\bx\by}^2)]=\frac{1}{2}\ln\left[2\pi
e\left(\hsigma_{\by}^2-\frac{\hC_{\bx\by}^2}{\hsigma_{\bx}^2}\right)\right],
\end{equation}
and so, the Gaussian empirical mutual information is given by
\begin{equation}
\hI_{\bx\by}(X;Y)=\hat{h}_{\by}(Y)-
\hat{h}_{\bx\by}(Y|X)=\frac{1}{2}\ln\frac{1}{1-\hat{\rho}_{\bx\by}^2}.
\end{equation}
Note that $\hat{h}_{\bx\by}(X|Y)$ can also be presented as
\begin{equation}
\hat{h}_{\bx\by}(X|Y)=\frac{1}{2}\ln[2\pi
e\hsigma_{\by|\bx}^2],
\end{equation}
where
\begin{equation}
\hsigma_{\by|\bx}^2=\min_{a\in\reals}\frac{1}{n}\|\by-a\bx\|^2=
\hsigma_{\by}^2(1-\hat{\rho}_{\bx\by}^2),
\end{equation}
and so, the notion of the Gaussian conditional empirical differential entropy
can easily apply to conditioning on more than one vector. For example,
given $(\bx,\bx^\prime,\by)\in (\reals^n)^3$,
\begin{equation}
\hat{h}_{\bx\bx^\prime\by}(Y|X,X^\prime)=\frac{1}{2}\ln[2\pi
e\hsigma_{\by|\bx\bx^\prime}^2],
\end{equation}
with
\begin{eqnarray}
\label{condempvar}
\hsigma_{\by|\bx\bx^\prime}^2&=&\min_{a,b\in\reals}\frac{1}{n}\|\by-
a\bx-b\bx^\prime\|^2\nonumber\\
&=&\min_{a,b\in\reals}\{\hsigma_{\by}^2-2(a\hC_{\bx\by}+b\hC_{\bx^\prime\by})+
(a^2+2ab\hat{\rho}_{\bx\bx^\prime}+b^2)P\}.
\end{eqnarray}
For future use, note that $\hsigma_{\by|\bx\bx^\prime}^2$
is a concave function of
$(\hsigma_{\by}^2,\hC_{\bx\by},\hC_{\bx^\prime\by})$, as it is obtained by
minimizing an affine function of these variables. Since the logarithmic
function is monotonically increasing and concave as well, then so is
$\hat{h}_{\bx\bx^\prime\by}(Y|X,X^\prime)$.
Finally, for a given $\epsilon > 0$
the Gaussian conditional type class of $\by\in\reals^n$ given
$\bx,\bx^\prime\in\reals^n$,
with tolerance $\epsilon$, will be defined as
\begin{equation}
\calT_{\epsilon}(\by|\bx,\bx^\prime)=
\left\{\tilde{\by}:~|\hsigma_{\tilde{\by}}^2-\hsigma_{\by}^2|\le\epsilon,~
|\hC_{\bx\tilde{\by}}-\hC_{\bx\by}|\le\epsilon,~
|\hC_{\bx^\prime\tilde{\by}}-\hC_{\bx^\prime\by}|\le\epsilon\right\}.
\end{equation}
In the sequel, we will need the following simple inequality:
\begin{equation}
\label{voltype}
\mbox{Vol}\{\calT_{\epsilon}(\by|\bx,\bx^\prime)\}\le
\exp\{n[\hat{h}_{\bx\bx^\prime\by}(Y|X,X^\prime)+O(\epsilon)]\},
\end{equation}
which is easily proved using the same technique as in \cite[Lemma 3]{me93}.

\subsection*{A.\ Proof of Theorem \ref{awgn-exponent}}

The proof is based on the same line of thought as in \cite[Theorem 1]{trc},
except that some modifications have to be carried out in order to address
the continuous alphabet case considered here. As in \cite{trc},
we use the identity
\begin{equation}
\bE\{\ln P_{\mbox{\tiny
e}}(\calC_n)\}=\lim_{r\to\infty}r\cdot\ln\left(\bE\{P_{\mbox{\tiny
e}}(\calC_n)^{1/r}\}\right),
\end{equation}
and so, we first focus on the calculation of $\bE\{P_{\mbox{\tiny
e}}(\calC_n)^{1/r}\}$. Another simple observation (easily proved using the
Chernoff bound) that will be used 
in the proof
is that for every $E > 0$, there exists a sufficiently large constant $B > 0$, 
such that for every $m\in\{0,1,\ldots,M-1\}$,
\begin{equation}
\mbox{Pr}\left\{\|\bY\|^2\ge nB~\mbox{or}~\|\bY\|^2\le n/B
\bigg|\bx[m]~\mbox{transmitted}\right\} \le e^{-nE},
\end{equation}
and so if we take $E =E_{\mbox{\tiny ex}}(0)$, the zero--rate expurgated
exponent, we can find a constant $B$
such that the contribution of $\mbox{Pr}\{\|\bY\|^2\le
n/B\}+\mbox{Pr}\{\|\bY\|^2\ge nB\}$ cannot affect
the error exponent $E_{\mbox{\tiny trc}}(R)\le E_{\mbox{\tiny ex}}(0)$ at any
coding rate (see also \cite{me93} for a similar argument).
In other words, if the decoder would declare an error for every $\by$ whose norm
is either larger than $nB$ or smaller than $n/B$, 
there would be no degradation in the error exponent.
Consequently, it is enough to focus 
only on $\by$--vectors whose norms are in the range
$[n/B,nB]$. Let us denote this set of vectors by $\calH_n(B)$. Now,
\begin{eqnarray}
P_{\mbox{\tiny e}}(\calC_n)&=&\frac{1}{M}\sum_{m=0}^{M-1}\sum_{\prm\ne
m}\int_{\reals^n}P(\by|\bx_m)\cdot\frac{\exp\{n\beta\bx^T[\prm]\cdot\by/\sigma^2\}}
{\sum_{\tm=0}^{M-1}
\exp\{n\beta\bx^T[\tm]\cdot\by/\sigma^2\}}\cdot\mbox{d}\by\nonumber\\
&=&\frac{1}{M}\sum_{m=0}^{M-1}\sum_{\prm\ne
m}\int_{\reals^n}P(\by|\bx_m)\cdot
\frac{\exp\{n\beta\bx^T[\prm]\cdot\by/\sigma^2\}}{\exp\{n\beta\bx^T[m]\cdot\by/\sigma^2\}
+\sum_{\tm\ne m}
\exp\{n\beta\bx^T[\tm]\cdot\by/\sigma^2\}}\cdot\mbox{d}\by\nonumber\\
&\exe&\frac{1}{M}\sum_{m=0}^{M-1}\sum_{\prm\ne
m}\int_{\calH_n(B)}P(\by|\bx_m)\cdot
\frac{\exp\{n\beta\bx^T[\prm]\cdot\by/\sigma^2\}}{\exp\{n\beta\bx^T[m]\cdot\by/\sigma^2\}
+Z_m(\by)}\cdot\mbox{d}\by,
\end{eqnarray}
where we have defined
\begin{equation}
Z_m(\by)=\sum_{\tm\ne m}
\exp\{n\beta\bx^T[\tm]\cdot\by/\sigma^2\}.
\end{equation}
We now argue that for every $\epsilon > 0$ (see Appendix for the proof),
\begin{equation}
\label{zmy}
\mbox{Pr}\left\{\calC_n:~\exists~0\le m <M,~\by\in\calH_n(B):~
Z_m(\by) <
\exp\{n[\alpha(R-\epsilon,\hsigma_{\by}^2)-\delta(\epsilon)]\}\right\}\exe 
e^{-n\infty},
\end{equation}
where $\delta(\epsilon)=2\beta\sqrt{P}B\epsilon/\sigma^2$.
We then have (neglecting $\epsilon$, which is arbitrarily small),
\begin{eqnarray}
\label{sheva}
\bE\left\{[P_{\mbox{\tiny e}}(\calC_n)]^{1/r}\right\}&\lexe&
\bE\left(\left[\frac{1}{M}\sum_{m=0}^{M-1}\sum_{\prm\ne
m}\int_{\calH_n(B)}P(\by|\bx_m)\times\right.\right.\nonumber\\
& &\left.\left.\min\left\{1,
\frac{\exp\{n\beta\bx^T[\prm]\by/\sigma^2\}}{\exp\{n\beta\bx^T[m]\by/\sigma^2\}
+\exp\{n\alpha(R,\hsigma_{\by}^2)\}}\right\}
\mbox{d}\by\right]^{1/r}\right)\nonumber\\
&\exe&\bE\left(\left[\frac{1}{M}\sum_{m=0}^{M-1}\sum_{\prm\ne
m}\int_{\calH_n(B)}P(\by|\bx_m)\times\right.\right.\nonumber\\
& &\left.\left.\min\left\{1,
\frac{\exp\{n\beta\bx^T[\prm]\by/\sigma^2\}}{\exp\{n\beta\bx^T[m]\by/\sigma^2\}
+\exp\{n\alpha(R,\hsigma_{\by}^2)
\}}\right\}\mbox{d}\by\right]^{1/r}\right)\nonumber\\
&\exe&\bE\left(\left[\frac{1}{M}\sum_{m=0}^{M-1}\sum_{\prm\ne
m}\int_{\calH_n(B)}P(\by|\bx_m)\times\right.\right.\nonumber\\
& &\left.\left.\exp\left\{-n\left[\max\left\{\frac{\beta\sqrt{P}
\hsigma_{\by}}{\sigma^2}\hat{\rho}_{\bx[m]\by},
\alpha(R,\hsigma_{\by}^2)\right\}
-\frac{\beta\sqrt{P}\hsigma_{\by}}{\sigma^2}\hat{\rho}_{\bx[\prm]\by}\right]_+\right\}
\mbox{d}\by\right]^{1/r}\right),\nonumber
\end{eqnarray}
where we have neglected the double--exponentially small contribution of
codes that violate (\ref{zmy}). The inner integral over $\by$ is of the form
$$\int_{\calH_n(B)}(2\pi\sigma^2)^{-n/2}\exp\left\{-\frac{\|\by-\bx[m]\|^2}{2\sigma^2}\right\}\cdot
\exp\{-nK(\hsigma_{\by},\hat{\rho}_{\bx[m]\by},\hat{\rho}_{\bx[\prm]\by})\}\mbox{d}\by,$$
where
$$K(\sigma_Y,\rho_{XY},\rho_{X^\prime Y})=\left[
\max\left\{\frac{\beta\sqrt{P}
\sigma_Y}{\sigma^2}\rho_{XY},
\alpha(R,\sigma_Y^2)\right\}
-\frac{\beta\sqrt{P}\sigma_Y}{\sigma^2}\rho_{X^\prime Y}\right]_+.$$
Since $K$ is not a quadratic function, this is not a simple Gaussian integral, but
its exponential order can be assessed using a Gaussian analogue of the
method of types (see, e.g., the analysis in \cite{me93}). For the
given two codewords, $\bx[m]$ and $\bx[\prm]$, we divide $\calH_n(B)$ into
conditional types $\{\calT_\epsilon(\by|\bx[m],\bx[\prm])\}$. The number of such
conditional types is finite: since $\|\by\|^2\le nB$ and
$\|\bx[m]\|^2=\|\bx[\prm]\|^2=nP$, then $\hC_{\bx[m]\by}$ and
$\hC_{\bx[\prm]\by}$ can
take on values only in the interval $[-\sqrt{PB},\sqrt{PB}]$, and so, there
are no more than $(B/\epsilon)\cdot(2\sqrt{PB}/\epsilon)^2=4PB^2/\epsilon^3$
conditional types classes $\{\calT_\epsilon(\by|\bx[m],\bx[\prm])\}$ within
$\calH_n(B)$, resulting from a grid of step size $\epsilon$ in each of the
ranges of $\hsigma_{\by}^2$, $\hC_{\bx[m]\by}$, and
$\hC_{\bx[\prm]\by}$. Therefore, the above integral is of the exponential order of
\begin{eqnarray}
& &\sup_{\by\in\calH_n(B)}\mbox{Vol}\{\calT_\epsilon(\by|\bx[m],\bx[\prm])\}\cdot
(2\pi\sigma^2)^{-n/2}\exp\left\{-\frac{\|\by-\bx[m]\|^2}{2\sigma^2}\right\}\times\nonumber\\
& &\exp\{-nK(\hsigma_{\by},\hat{\rho}_{\bx[m]\by},\hat{\rho}_{\bx[\prm]\by})\}\nonumber\\
&\exe&\exp\left\{-n\inf_{\by}\left[\frac{1}{2}\ln(2\pi\sigma^2)+
\frac{\hsigma_{\by}^2-2\sqrt{P}\hsigma_{\by}+P}{2\sigma^2}-
\hat{h}_{\by|\bx[m]\bx[\prm]}(Y|X,X^\prime)+\right.\right.\nonumber\\
& &\left.\left.K(\hsigma_{\by},
\hat{\rho}_{\bx[m]\by},\hat{\rho}_{\bx[\prm]\by})\right]\right\}\nonumber\\
&\lexe&\exp\left\{-n\inf_{\sigma_Y,\rho_{XY},\rho_{X^\prime Y}}\left[
\frac{1}{2}\ln(2\pi\sigma^2)+\frac{\sigma_Y^2-2\sqrt{P}\sigma_Y+P}{2\sigma^2}-
\frac{1}{2}\ln(2\pi
e\sigma_{Y|XX^\prime}^2)+\right.\right.\nonumber\\
& &\left.\left.K(\sigma_Y,\rho_{XY},\rho_{X^\prime
Y})\right]\right\},
\end{eqnarray}
where the infimum over $(\sigma_Y,\rho_{XY},\rho_{X^\prime Y})$ is such that
the matrix (\ref{covmat}) (with $\rho_{XX^\prime}=\hat{\rho}_{\bx[m]\bx[\prm]}$) 
is positive semi--definite.
Using eqs.\ (\ref{condempvar}) and (\ref{voltype}), 
it is readily observed that this integral is of
the exponential order of $e^{-n\Gamma(\rho_{\bx[m]\bx[\prm]})}$, where the
function $\Gamma(\cdot)$ is defined in (\ref{Gammadef}).
Thus, we have shown that
\begin{eqnarray}
\bE\left\{[P_{\mbox{\tiny e}}(\calC_n)]^{1/r}\right\}&\lexe&
\bE\left(\left[\frac{1}{M}\sum_{m=0}^{M-1}\sum_{\prm\ne
m}\exp\{-n\Gamma(\rho_{\bx[m]\bx[\prm]})\}\right]^{1/r}\right)\nonumber\\
&\le&e^{nR/r}\bE\left(\sum_{\rho_i}M(\rho_i)\exp\{-n\Gamma(\rho_i)\}\right)^{1/r}\nonumber\\
&\le&e^{nR/r}\sum_{\rho_i}\bE\{M(\rho_i)^{1/r}\}\cdot\exp\{-n\Gamma(\rho_i)/r\}
\end{eqnarray}
where $\{\rho_i\}$ form a fine quantization grid over the interval $(-1,1)$
and $M(\rho_i)$ is the number of codeword pairs $\{\bx[m],\bx[\prm]\}$ whose
empirical correlation coefficient fall in the quantization bin of $\rho_i$.
The remaining part of the proof follows exactly the same lines as the proof of
\cite[Theorem 1]{trc}, except that joint types
$\{Q_{XX^\prime}\}$ (and the conditional types, $\{Q_{X^\prime|X}\}$) of \cite{trc}
are now indexed by $\{\rho_i\}$ and $I(Q)$ is replaced by
$\frac{1}{2}\ln\frac{1}{1-\rho_{XX^\prime}^2}$.

\subsection*{B.\ Proof of Theorem \ref{thm2}}

The proof goes along the same lines as the proof of Theorem
\ref{awgn-exponent}, except
that there are $n$ independent copies of all elements produced from
$\ell$--vectors, and so, we will outline only the main differences.
Here we denote
\begin{eqnarray}
\hsigma_{\by_i}^2&=&\frac{\|\by_i\|^2}{\ell}\\
\hC_{\bx_i\by_i}&=&\frac{\bx_i^T\by_i}{\ell}.
\end{eqnarray}
For a given code $\calC_n$, a given transmitted codeword $\bx[m]$
and a given competing codeword $\bx[\prm]$,
\begin{eqnarray}
& &\mbox{Pr}\{\prm~\mbox{decoded}~\bigg|~m~\mbox{transmitted}\}\nonumber\\
&=&\int_{\reals^N}P(\by|\bx[m])\cdot
\mbox{Pr}\{\hat{m}=\prm|\by\}\mbox{d}\by\nonumber\\
&\le&\int_{\reals^N}\left[\prod_{i=0}^{n-1}(2\pi\sigma_i^2)^{-\ell/2}
\exp\{-\|\by_i-\bx_i[m]\|^2/(2\sigma_i^2)\}\right]\times\nonumber\\
& &\min\left\{1,\exp\left[\beta\sum_{i=0}^{n-1}\left(\frac{\bx_i^T[\prm]\by_i}{\tsigma_i^2}
-\frac{\bx_i^T[m]\by_i}{\tsigma_i^2}\right)\right]\right\}\mbox{d}\by\nonumber\\
&\exe&\max_{\{\hsigma_{\by_i}^2,\hC_{\bx_i[m]\by},\hC_{\bx_i[\prm]\by}\}_i}
\left[\prod_{i=0}^{n-1}\mbox{Vol}
\{\calT_\epsilon(\by_i|\bx_i[m],\bx_i[\prm])\}\times\right.\nonumber\\
& &\left.(2\pi\sigma_i^2)^{-\ell/2}
\exp\{-\|\by_i-\bx_i[m]\|^2/(2\sigma_i^2)\}\right]\cdot
\exp\left\{-\beta\left[\sum_{i=0}^{n-1}\left(\frac{\bx_i^T[m]\by_i}{\tsigma_i^2}-
\frac{\bx_i^T[\prm]\by_i}{\tsigma_i^2}\right)\right]_+\right\}\nonumber\\
&\exe&\exp\left\{-\ell\min_{\{\hsigma_{\by_i}^2,\hC_{\bx_i[m]\by},
\hC_{\bx_i[\prm]\by}\}_i}\sum_{i=0}^{n-1}\left(\frac{1}{2}\ln(2\pi\sigma_i^2)+
\frac{\hsigma_{\by_i}^2-2\hC_{\bx_i[m]\by_i}+P_i}{2\sigma_i^2}-\right.\right.\nonumber\\
& &\left.\left.\hat{h}_{\bx_i[m]\bx_i[\prm]\by}(Y|X,X^\prime)\right)+
\beta\left[\sum_{i=0}^{n-1}
\frac{\hC_{\bx_i[m]\by_i}-\hC_{\bx_i[\prm]\by_i}}{\tsigma_i^2}\right]_+\right\}\nonumber\\
&\exe&\exp\left\{-\ell\inf_{\{\sigma_{\by_i}^2,\hC_{\bx_i[m]\by_i},
\hC_{\bx_i[\prm]\by_i}\}_i}\sum_{i=0}^{n-1}\left(\frac{1}{2}\ln(2\pi\sigma_i^2)+
\frac{\hsigma_{\by_i}^2-2\hC_{\bx_i[m]\by_i}+P_i}{2\sigma_i^2}-\right.\right.\nonumber\\
& &\left.\left.\hat{h}_{\bx_i[m]\bx_i[\prm]\by}(Y|X,X^\prime)\right)+
\sup_{0\le\lambda\le\beta}\sum_{i=0}^{n-1}
\frac{\hC_{\bx_i[m]\by_i}-\hC_{\bx_i[\prm]\by_i}}{\tsigma_i^2}\right\}\nonumber\\
&=&\exp\left\{-\ell\sup_{0\le\lambda\le\beta}
\inf_{\{\sigma_{\by_i}^2,\hC_{\bx_i[m]\by_i},
\hC_{\bx_i[\prm]\by_i}\}_i}\sum_{i=0}^{n-1}\left(\frac{1}{2}\ln(2\pi\sigma_i^2)+
\frac{\hsigma_{\by_i}^2-2\hC_{\bx_i[m]\by_i}+P_i}{2\sigma_i^2}-\right.\right.\nonumber\\
& &\left.\left.\hat{h}_{\bx_i[m]\bx_i[\prm]\by}(Y|X,X^\prime)\right)+
\sum_{i=0}^{n-1}
\frac{\hC_{\bx_i[m]\by_i}-\hC_{\bx_i[\prm]\by_i}}{\tsigma_i^2}\right\}\nonumber\\
&=&\exp\left\{-\ell\sup_{0\le\lambda\le\beta}\sum_{i=0}^{n-1}
\inf_{\{\sigma_{\by_i}^2,\hC_{\bx_i[m]\by_i},
\hC_{\bx_i[\prm]\by_i}\}_i}\left(\frac{1}{2}\ln(2\pi\sigma_i^2)+
\frac{\hsigma_{\by_i}^2-2\hC_{\bx_i[m]\by_i}+P_i}{2\sigma_i^2}-\right.\right.\nonumber\\
& &\left.\left.\hat{h}_{\bx_i[m]\bx_i[\prm]\by}(Y|X,X^\prime)+
\sum_{i=0}^{n-1}
\frac{\hC_{\bx_i[m]\by_i}-\hC_{\bx_i[\prm]\by_i}}{\tsigma_i^2}\right)\right\}\nonumber\\
&=&\exp\left\{-\ell\sup_{0\le\lambda\le\beta}\sum_{i=0}^{n-1}\lambda\mu_i(1-\lambda\mu_i)
\mbox{snr}_i(1-\rho_i[m,\prm])\right\},
\end{eqnarray}
where $\rho_i[m,\prm]$ is the empirical correlation coefficient between
$\bx_i[m]$ and $\bx_i[\prm]$, and where the last step is obtained from the
minimization of each term over $\{\sigma_{\by_i}^2,\hC_{\bx_i[m]\by_i},
\hC_{\bx_i[\prm]\by_i}\}$, very similarly as is done in eq.\
(\ref{GammaLoptimization}).
The proof is completed similarly as in \cite{trc} by showing that the typical
number of codeword pairs $\{\bx[m],\bx[\prm]\}$ for which the segmental
empirical correlation coefficients are around
$(\rho_0,\rho_1,\ldots,\rho_{n-1})$ (within a fine grid) is
of the exponential order of
\begin{equation}
M(\rho_0,\rho_1,\ldots,\rho_{n-1})\exe
\exp\left\{N\left[2R-\frac{1}{2n}
\sum_{i=0}^{n-1}\ln\frac{1}{1-\rho_i^2}\right]\right\}
\end{equation}
as long as $\frac{1}{2n}\sum_{i=0}^{n-1}\ln\frac{1}{1-\rho_i^2} < 2R$, and
$M(\rho_0,\rho_1,\ldots,\rho_{n-1})=0$ otherwise.
Thus, for the typical code,
\begin{eqnarray}
P_{\mbox{\tiny e}}(\calC_n)&\lexe&
\frac{1}{M}\sum_{m=0}^{M-1}\sum_{\prm\ne m}
\exp\left\{-\ell\sup_{0\le\lambda\le\beta}\sum_{i=0}^{n-1}\lambda\mu_i(1-\lambda\mu_i)
\mbox{snr}_i(1-\rho_i[m,\prm])\right\}\nonumber\\
&\exe&
e^{-NR}\sum_{\rho_0,\rho_1,\ldots,\rho_{n-1}}M(\rho_0,\rho_1,\ldots,\rho_{n-1})\times\nonumber\\
& &\exp\left\{-\ell\sup_{0\le\lambda\le\beta}\sum_{i=0}^{n-1}\lambda\mu_i(1-\lambda\mu_i)
\mbox{snr}_i(1-\rho_i)\right\}\nonumber\\
&\exe&\exp\left\{-N\inf_{\rho_0,\rho_1,\ldots,\rho_{n-1}}\sup_{0\le\lambda\le\beta}
\frac{1}{n}\sum_{i=0}^{n-1}\left[\lambda\mu_i(1-\lambda\mu_i)\mbox{snr}_i(1-\rho_i)+
\frac{1}{2}\ln\frac{1}{1-\rho_i^2}\right]-R\right\}\nonumber\\
&\exe&\exp\left\{-N\sup_{0\le\lambda\le\beta}\inf_{\rho_0,\rho_1,\ldots,\rho_{n-1}}
\frac{1}{n}\sum_{i=0}^{n-1}\left[\lambda\mu_i(1-\lambda\mu_i)\mbox{snr}_i(1-\rho_i)+
\frac{1}{2}\ln\frac{1}{1-\rho_i^2}\right]-R\right\},\nonumber
\end{eqnarray}
where the infimum is subject to the constraint
$\frac{1}{2n}\sum_{i=0}^{n-1}\ln\frac{1}{1-\rho_i^2} < 2R$.
This constrained minimization at the exponent can be presented as
\begin{eqnarray}
& &\inf_{\rho_0,\rho_1,\ldots,\rho_{n-1}}\sup_{\vartheta\ge 0}
\frac{1}{n}\sum_{i=0}^{n-1}\left[\lambda\mu_i(1-\lambda\mu_i)\mbox{snr}_i(1-\rho_i)+
\frac{1}{2}\ln\frac{1}{1-\rho_i^2}+
\vartheta\left(\frac{1}{2}\ln\frac{1}{1-\rho_i^2}-2R\right)\right]-R\nonumber\\
&=&\sup_{\vartheta\ge 0}\frac{1}{n}\sum_{i=0}^{n-1}\inf_{\rho_i}\left[
\lambda\mu_i(1-\lambda\mu_i)\mbox{snr}_i(1-\rho_i)+
\frac{1+\vartheta}{2}\ln\frac{1}{1-\rho_i^2}\right]
-(2\vartheta+1)R\nonumber\\
&=&\sup_{\theta\ge 1}\frac{1}{n}\sum_{i=0}^{n-1}\inf_{\rho_i}
\left[\lambda\mu_i(1-\lambda\mu_i)\mbox{snr}_i(1-\rho_i)+
\frac{\theta}{2}\ln\frac{1}{1-\rho_i^2}\right]
-(2\theta-1)R\nonumber\\
&=&\sup_{\theta\ge
1}\frac{1}{n}\sum_{i=0}^{n-1}A(\mbox{snr}_i,\mu_i,\lambda,\theta)-(2\theta-1)R,
\end{eqnarray}
which is the relevant expression for Theorem \ref{thm2}.

\section*{Appendix A: Proof of Eq.\ (\ref{zmy})}
\renewcommand{\theequation}{A.\arabic{equation}}
    \setcounter{equation}{0}

For a given $\epsilon > 0$,
consider the set of vectors,
\begin{equation}
H_n^\epsilon(B)=H_n(B)\bigcap\left\{\by=(i_1,i_2,\ldots,i_n)\cdot 2\epsilon:~
(i_1,i_2,\ldots,i_n)\in\calZ^n\right\},
\end{equation}
namely, the grid of all vectors within $H_n(B)$ whose components are
integer multiples of $2\epsilon$. Obviously,
\begin{equation}
|H_n^\epsilon(B)|\le\frac{\mbox{Vol}\{H_n(B)\}}{(2\epsilon)^n}\le
\frac{(2\pi e B)^{n/2}}{(2\epsilon)^n}=\left(\frac{\sqrt{2\pi e
B}}{2\epsilon}\right)^n,
\end{equation}
in other words, the number of points within $H_n^\epsilon(B)$ is
exponential in $n$. If we prove that
\begin{equation}
\label{zmy1}
\mbox{Pr}\left\{\exists m\in\{0,1,\ldots,M-1\},~\by\in\calH_n^\epsilon(B):~
Z_m(\by) < \exp\{n\alpha(R-\epsilon,\hsigma_{\by}^2)\right\}\exe e^{-n\infty}.
\end{equation}
then the result will follow by 
continuity considerations as follows: if $\by$ and $\by^\prime$ differ by no
more than $\epsilon$ component-wise in absolute value, then
for $\epsilon \ll 1/\sqrt{B}$,
\begin{eqnarray}
\hsigma_{\by}&=&\sqrt{\frac{1}{n}\|\by\|^2}\nonumber\\
&\ge&\sqrt{\frac{1}{n}(\|\by^\prime\|^2-2\sqrt{Bn}\sqrt{n\epsilon^2}+n\epsilon^2)}\nonumber\\
&\ge&\sqrt{\hsigma_{\by^\prime}^2-2\sqrt{B}\epsilon}\nonumber\\
&\ge&\hsigma_{\by^\prime}-\frac{2\sqrt{B}\epsilon}{2\hsigma_{\by^\prime}}\nonumber\\
&\ge&\hsigma_{\by^\prime}-\frac{2\sqrt{B}\epsilon}{2/\sqrt{B}}\nonumber\\
&=&\hsigma_{\by^\prime}-B\epsilon.
\end{eqnarray}
Thus,
\begin{eqnarray}
\alpha(R,\hsigma_{\by}^2)&=&
\sup_{|\rho|\le\sqrt{1-e^{-2R}}}\left[\frac{\beta\sqrt{P}\hsigma_{\by}\rho}
{\sigma^2}+\frac{1}{2}\ln(1-\rho^2)\right]+R\nonumber\\
&\ge&\sup_{|\rho|\le\sqrt{1-e^{-2R}}}\left[\frac{\beta\sqrt{P}
(\hsigma_{\by^\prime}-B\epsilon)\rho}
{\sigma^2}+\frac{1}{2}\ln(1-\rho^2)\right]+R\nonumber\\
&\ge&\alpha(R,\hsigma_{\by^\prime}^2)-\frac{\beta\sqrt{P}B\epsilon}{\sigma^2}.
\end{eqnarray}
On the other hand,
\begin{eqnarray}
Z_m(\by)&=&\sum_{m^\prime\ne
m}\exp\left\{\frac{\beta\bx^T[m^\prime]\by}{\sigma^2}\right\}\nonumber\\
&\le&\sum_{m^\prime\ne
m}\exp\left\{\frac{\beta(\bx^T[m^\prime]\by^\prime+
\sqrt{nP}\sqrt{n\epsilon^2})}{\sigma^2}\right\}\nonumber\\
&=&\exp\left\{\frac{n\beta\sqrt{P}\epsilon}{\sigma^2}\right\}\cdot
Z_m(\by^\prime).
\end{eqnarray}
It follows that if $Z_m(\by)\ge
\exp\{n\alpha(R-\epsilon,\hsigma_{\by}^2)$
for all $(m,\by)\in \{0,1,\ldots,M-1\}\times
\calH_n^\epsilon(B)$, then
$Z_m(\by)\ge
\exp\{n[\alpha(R-\epsilon,\hsigma_{\by}^2)-\delta(\epsilon)]\}$
for all $(m,\by)\in \{0,1,\ldots,M-1\}\times
\calH_n(B)$, where $\delta(\epsilon)$ is as defined above.

Since the number of pairs $(m,\by)\in\{0,1,\ldots,M-1\}\times \calH_n^\epsilon(B)$ is
upper bounded by $e^{nR}\cdot(\sqrt{2\pi e
B}/\epsilon)^n$, which is exponential in $n$, it is enough to prove that
\begin{equation}
\label{zmy2}
\mbox{Pr}\left\{
Z_m(\by) < \exp\{n\alpha(R-\epsilon,\hsigma_{\by}^2)\right\}\exe e^{-n\infty}
\end{equation}
for every given
$(m,\by)\in\{0,1,\ldots,M-1\}\times\calH_n^\epsilon(B)$, as the union
bound over all these pairs will not affect the super--exponential decay of the
probability under discussion. The proof of this fact is the same as the
proof of the analogous result in \cite[Appendix B]{gld}, except that here
$Z_m(\by)$ is approximated as $\sum_i M(\rho_i)\exp\{\beta
n\sqrt{P}\hsigma_{\by}\rho_i/\sigma^2\}$, where $\{\rho_i\}$ form a fine
quantization grid with spacing $\epsilon$ within
the interval $(-1,1)$ and $M(\rho_i)$ is the number of codewords other than
$\bx[m]$ whose empirical correlation coefficient with $\by$ is
between $\rho_i-\epsilon/2$ and $\rho_i+\epsilon/2$. The proof in \cite[Appendix
B]{gld} applies here verbatim except that $Q$, $g(Q)$ and $I(Q)$ of \cite{gld}
are replaced by
$\rho_i$, $\beta\sqrt{P}\hsigma_{\by}\rho_i$ and
$\frac{1}{2}\ln\frac{1}{1-\rho_i^2}$, respectively.

\section*{Appendix B: Calculation of $R_0$}
\renewcommand{\theequation}{B.\arabic{equation}}
    \setcounter{equation}{0}

In ordinary random coding under the same regime, we have the following.
The pairwise error event is:
\begin{equation}
\sum_{i=0}^{n-1}\frac{\tilde{\bx}_i^T\by_i}{\tsigma_i^2}\ge
\sum_{i=0}^{n-1}\frac{\bx_i^T\by_i}{\tsigma_i^2},
\end{equation}
where $\bx_i$ represents the $i$--th segment of the transmitted codeword and
$\tilde{\bx}_i$ stands for that of a competing one.
On substituting $\by_i=\bx_i+\bz_i$, this becomes
\begin{equation}
\sum_{i=0}^{n-1}\frac{\tilde{\bx}_i^T(\bx_i+\bz_i)}{\tsigma_i^2}\ge
\sum_{i=0}^{n-1}\frac{\bx_i^T(\bx_i+\bz_i)}{\tsigma_i^2},
\end{equation}
or
\begin{equation}
\sum_{i=0}^{n-1}\frac{(\tilde{\bx}_i-\bx_i)^T\bz_i}{\tsigma_i^2}\ge
\ell\sum_{i=0}^{n-1}\frac{P_i(1-\rho_i)}{\tsigma_i^2},
\end{equation}
where $\rho_i$ is the empirical correlation coefficient between
$\bx_i$ and $\tilde{\bx}_i$.
Now,
\begin{equation}
\sum_{i=0}^{n-1}\frac{(\tilde{\bx}_i-\bx_i)^T\bz_i}{\tsigma_i^2}
\sim\calN\left(0,2\ell\sum_{i=0}^{n-1}\frac{P_i\sigma_i^2
(1-\rho_i)}{\tsigma_i^4}\right),
\end{equation}
thus, the probability of the above event is
\begin{eqnarray}
\mbox{Pr}\left\{\sum_{i=0}^{n-1}\frac{(\tilde{\bx}_i-\bx_i)^T\bz_i}{\tsigma_i^2}\ge
\ell\sum_{i=0}^{n-1}\frac{P_i(1-\rho_i)}{\tsigma_i^2}\right\}
&=&Q\left(\frac{\ell\sum_{i=0}^{n-1}P_i(1-\rho_i)/\tsigma_i^2}
{\sqrt{2\ell\sum_{i=0}^{n-1}P_i\sigma_i^2(1-\rho_i)/\tsigma_i^4}}\right)\nonumber\\
&\exe&\exp\left\{-\frac{\ell\left[\sum_{i=0}^{n-1}P_i(1-\rho_i)/\tsigma_i^2\right]^2}
{4\sum_{i=0}^{n-1}P_i\sigma_i^2(1-\rho_i)/\tsigma_i^4}\right\},
\end{eqnarray}
where $Q(\cdot)$ is the error function,
\begin{equation}
Q(x)=\frac{1}{\sqrt{2\pi}}\int_x^\infty e^{-u^2/2}\mbox{d}u.
\end{equation}
The average pairwise probability of error is therefore given by
\begin{eqnarray}
& &\overline{\mbox{Pr}}\left\{\sum_{i=0}^{n-1}
\frac{(\tilde{\bx}_i-\bx_i)^T\bz_i}{\tsigma_i^2}\ge
\ell\sum_{i=0}^{n-1}\frac{P_i(1-\rho_i)}{\tsigma_i^2}\right\}\nonumber\\
&\exe&\max_{\rho_1,\ldots,\rho_n}
\exp\left\{\ell\sum_{i=0}^{n-1}\frac{1}{2}\ln(1-\rho_i^2)-
\frac{\ell\left[\sum_{i=0}^{n-1}P_i(1-\rho_i)/\tsigma_i^2\right]^2}
{4\sum_{i=0}^{n-1}P_i\sigma_i^2(1-\rho_i)/\tsigma_i^4}\right\},
\end{eqnarray}
where the term $\ell\sum_{i=0}^{n-1}\frac{1}{2}\ln(1-\rho_i^2)$ accounts for
the probabilistic weight of $(\rho_0,\ldots,\rho_{n-1})$.
Therefore, the exponent is
\begin{eqnarray}
& &\min_{\rho_0,\ldots,\rho_{n-1}}
\left\{\frac{1}{n}\sum_{i=0}^{n-1}\frac{1}{2}\ln\frac{1}{1-\rho_i^2}+
\frac{\left[\sum_{i=0}^{n-1}P_i(1-\rho_i)/\tsigma_i^2\right]^2}
{4n\sum_{i=0}^{n-1}P_i\sigma_i^2(1-\rho_i)/\tsigma_i^4}\right\}\nonumber\\
&=&\min_{\rho_0,\ldots,\rho_{n-1}}\left\{-\frac{1}{n}\sum_{i=0}^{n-1}\frac{1}{2}\ln(1-\rho_i^2)+
\sup_{\lambda\ge
0}\left[\lambda\cdot\frac{1}{n}\sum_{i=0}^{n-1}\frac{P_i(1-\rho_i)}{\tsigma_i^2}-\lambda^2\cdot
\frac{1}{n}\sum_{i=0}^{n-1}\frac{P_i\sigma_i^2(1-\rho_i)}{\tsigma_i^4}\right]\right\}\nonumber\\
&=&\sup_{\lambda\ge
0}\frac{1}{n}\sum_{i=0}^{n-1}\min_{\rho_i}\left\{\frac{1}{2}\ln\frac{1}{1-\rho_i^2}
+\lambda\cdot\frac{P_i(1-\rho_i)}{\tsigma_i^2}-\lambda^2
\cdot\frac{P_i\sigma_i^2(1-\rho_i)}{\tsigma_i^4}\right\}\nonumber\\
&=&\sup_{\lambda\ge
0}\frac{1}{n}\sum_{i=0}^{n-1}\min_{\rho_i}\left\{\frac{1}{2}\ln\frac{1}{1-\rho_i^2}
+\frac{P_i}{\tsigma_i^2}\left(\lambda-\frac{\lambda^2\sigma_i^2}{\tsigma_i^2}\right)
(1-\rho_i)\right\}\nonumber\\
&=&\sup_{\lambda\ge
0}\frac{1}{n}\sum_{i=0}^{n-1}A(\mbox{snr}_i,\mu_i,\lambda,1),\nonumber\\
\end{eqnarray}
and in the limit of large $n$,
\begin{equation}
R_0=\sup_{\lambda\ge
0}\frac{1}{2\pi}\int_{0}^{2\pi}A[\mbox{snr}(\omega),\mu(\omega),\lambda,1]\mbox{d}\omega=B(1),
\end{equation}
as opposed to the zero--rate TRC exponent, which corresponds to
$$\sup_{\lambda\ge
0}\frac{1}{2\pi}\int_{0}^{2\pi}
A[\mbox{snr}(\omega),\mu(\omega),\lambda,\infty]\mbox{d}\omega.$$

\section*{Appendix C: Tightness at Low Rates}
\renewcommand{\theequation}{C.\arabic{equation}}
    \setcounter{equation}{0}

As before, we begin from parallel channels, and then take the limit $n\to\infty$
in order to pass to the continuous frequency domain. 
For deterministic decoding we also take the limit $\beta\to\infty$. Let us denote 
\begin{equation}
G(\bx,\by)=\frac{1}{n}\sum_{i=0}^{n-1}\frac{\bx_i^T\by_i}{\tsigma_i^2}
\end{equation}
and
\begin{equation}
\alpha(R,\by)=\sup\sum_{i=0}^{n-1}
\frac{\sqrt{P_i}\hsigma_{\by_i}\rho_i}{\tsigma_i^2},
\end{equation}
where the supermum is over all $(\rho_0,\rho_1,\ldots,\rho_{n-1})$ such that
$\sum_{i=0}^{n-1}\frac{1}{2}\ln\frac{1}{1-\rho_i^2}\le nR$.
As in \cite{trc}, the TRC exponent for deterministic mismatched decoding is obtained by
analyzing the probability of the event
$G(\bx_{m^\prime},\by)\ge\max\{G(\bx_m,\by),\alpha(R,\by)\}$, for a given
code, taking the logarithm, and finally, averaging over the code ensemble.
As mentioned earlier, here we removed
the term $\alpha(\by,R)$ and upper bounded by this probability by the
probability of the event $G(\bx_{m^\prime},\by)\ge G(\bx_m,\by)$, which is the
union of pairwise error events. We would like to show now that there is a
range of low rates $[0,R_{\mbox{\tiny t}}]$, 
where this bound is exponentially tight. Consider the
chain of inequalities
\begin{eqnarray}
& &\mbox{Pr}\{G(\bx[m^\prime],\by)\ge G(\bx[m],\by)\}\nonumber\\
&=&\mbox{Pr}\{G(\bx[m^\prime],\by)\ge G(\bx[m],\by)\ge\alpha(R,\by)\}+
\mbox{Pr}\{G(\bx[m^\prime],\by)\ge \alpha(R,\by)\ge
G(\bx[m],\by)\}+\nonumber\\
& &\mbox{Pr}\{\alpha(R,\by)\ge G(\bx[m^\prime],\by)\ge G(\bx[m],\by)\}\nonumber\\
&=&\mbox{Pr}\{G(\bx[m^\prime],\by)\ge \max\{\alpha(R,\by),G(\bx[m],\by)\}+
\mbox{Pr}\{\alpha(R,\by)\ge G(\bx[m^\prime],\by)\ge G(\bx[m],\by)\}\nonumber\\
&\le&\mbox{Pr}\{G(\bx[m^\prime],\by)\ge \max\{\alpha(R,\by),G(\bx[m],\by)\}+
\mbox{Pr}\{\alpha(R,\by)\ge G(\bx[m],\by)\},
\end{eqnarray}
where the left--most side is the quantity we analyze in the proof of Theorem 2, 
and in the right--most
side, the first term is the desired quantity, whereas the second term is the
supplementary term that we would now like to focus on. In particular, if we
show that the second term decays at an exponential rate faster than that of the first
term (which is $E_{\mbox{\tiny trc}}^-(R)$), then our union--bound analysis in
Theorem 2 is tight. Now,
\begin{equation}
\mbox{Pr}\{\alpha(R,\by)\ge G(\bx[m],\by)\}=
\mbox{Pr}\left\{\max_{\{-\sum_{i=0}^{n-1}\ln(1-\hat{\rho}_i^2)\le
2nR\}}\sum_{i=0}^{n-1}
\frac{\sqrt{P_i}
\hsigma_{\by_i}\hat{\rho}_i}{\tsigma_i^2}\ge
\frac{1}{\ell}\sum_{i=0}^{n-1}\frac{\bx_i^T[m]\by_i}{\tsigma_i^2}\right\}.
\end{equation}
But
\begin{eqnarray}
& &\max_{\{-\sum_{i=0}^{n-1}\ln(1-\hat{\rho}_i^2)\le 2nR\}}\sum_i
\frac{\sqrt{P_i}
\hsigma_{\by_i}\hat{\rho}_i}{\tsigma_i^2}\nonumber\\
&=&\max_{\hat{\rho}_0,\ldots,\hat{\rho}_{n-1}}\inf_{\lambda\ge 0}
\sum_{i=0}^{n-1}\left\{\frac{\sqrt{P_i}\hsigma_{\by_i}
\hat{\rho}_i}{\tsigma_i^2}+\lambda\left[R+
\frac{1}{2}\ln(1-\hat{\rho}_i^2)\right]\right\}\nonumber\\
&=&\inf_{\lambda\ge 0}
\sum_{0=1}^{n-1}\max_{\rho_i}\left\{\frac{\sqrt{P_i}
\hsigma_{\by_i}\hat{\rho}_i}{\tsigma_i^2}+\lambda\left[R+
\frac{1}{2}\ln(1-\hat{\rho}_i^2)\right]\right\}\nonumber\\
&=&\inf_{\lambda\ge 0}\left\{
\sum_{0=1}^{n-1}W\left(\frac{\sqrt{P_i}\hsigma_{\by_i}}{\tsigma_i^2},\lambda\right)+\lambda
R\right\},
\end{eqnarray}
where we define
\begin{eqnarray}
W(\alpha,\lambda)&=&\max_{\rho}
\left\{\alpha\rho+\frac{\lambda}{2}\ln(1-\rho^2)\right\}\nonumber\\
&=&\lambda\cdot w\left(\frac{\alpha}{\lambda},\infty\right)\nonumber\\
&=&\frac{2\alpha^2}{\lambda+\sqrt{\lambda^2+4\alpha^2}}+
\frac{\lambda}{2}\ln\left[\frac{2\lambda}{\lambda+\sqrt{\lambda^2+4\alpha^2}}\right].
\end{eqnarray}
Now,
\begin{eqnarray}
\mbox{Pr}\{\alpha(R,\by)\ge G(\bx[m],\by)\}&=&
\mbox{Pr}\left\{\frac{1}{\ell}\sum_{i=0}^{n-1}\frac{\bx_i^T[m]\by_i}{\tsigma_i^2}\le
\inf_{\lambda\ge 0}\left[
\sum_{i=1}^kW\left(\frac{\sqrt{P_i}\hsigma_{\by_i}}{\tsigma_i^2},\lambda\right)+\lambda
R\right]\right\}\nonumber\\
&\le&\inf_{\lambda\ge 0}
\mbox{Pr}\left\{\frac{1}{\ell}\sum_{i=0}^{n-1}\frac{\bx_i^T[m]\by_i}{\tsigma_i^2}\le
\sum_{i=0}^{n-1}\left[W\left(\frac{\sqrt{P_i}\hsigma_{\by_i}
}{\tsigma_i^2},\lambda\right)+\lambda
R\right]\right\}.
\end{eqnarray}
For a given $\lambda\ge 0$, the exponent of the last probability is given by
\begin{equation}
\min_{\{\rho_i,\hsigma_{\by_i}^2\}}\frac{1}{n}\sum_{i=0}^{n-1}\left\{
\frac{1}{2}\ln(2\pi\sigma_i^2)+\frac{\hsigma_{\by_i}^2-2\sqrt{P_i}
\hsigma_{\by_i}\rho_i+P_i}{2\sigma_i^2}
-\frac{1}{2}\ln[2\pi e\hsigma_{\by_i}^2(1-\rho_i^2)]\right\},
\end{equation}
where the minimum is taken over 
all $\{(\rho_i,\hsigma_{\by_i}^2),~0=1,2,\ldots,n-1\}$ such that
\begin{equation}
\sum_{i=0}^{n-1}\frac{\sqrt{P_i}\hsigma_{\by_i}\rho_i}{\tsigma_i^2}\le
\sum_{i=0}^{n-1}\left[W\left(\frac{\sqrt{P_i}
\hsigma_{\by_i}}{\tsigma_i^2},\lambda\right)+\lambda
R\right],
\end{equation}
or, equivalently, the exponent is given by
\begin{eqnarray}
& &\min_{\{\rho_i,\hsigma_{\by_i}^2\}}\sup_{\zeta\ge 0}
\frac{1}{n}\sum_{0=1}^{n-1}\left\{
\frac{1}{2}\ln(2\pi\sigma_i^2)+\frac{\hsigma_{\by_i}^2-2\sqrt{P_i}\hsigma_{\by_i}
\rho_i+P_i}{2\sigma_i^2}
-\frac{1}{2}\ln[2\pi
e\hsigma_{\by_i}^2(1-\rho_i^2)]+\right.\nonumber\\
& &\left.\frac{\zeta}{\lambda}
\left[\frac{\sqrt{P_i}\hsigma_{\by_i}\rho_i}{\tsigma_i^2}-W\left(\frac{\sqrt{P_i}
\hsigma_{\by_i}}{\tsigma_i^2},\lambda\right)-\lambda R\right]\right\}\nonumber\\
&\ge&\sup_{\zeta\ge 0}
\frac{1}{n}\sum_{i=0}^{n-1}\min_{\rho_i,\hsigma_{\by_i}^2}\left\{
\frac{1}{2}\ln(2\pi\sigma_i^2)+\frac{\hsigma_{\by_i}^2-2\sqrt{P_i}\hsigma_{\by_i}
\rho_i+P_i}{2\sigma_i^2}
-\frac{1}{2}\ln[2\pi
e\hsigma_{\by_i}^2(1-\rho_i^2)]+\right.\nonumber\\
& &\left.\frac{\zeta}{\lambda}
\left[\frac{\sqrt{P_i}\hsigma_{\by_i}\rho_i}{\tsigma_i^2}-W\left(\frac{\sqrt{P_i}
\hsigma_{\by_i}}{\tsigma_i^2},\lambda\right)-\lambda R\right]\right\}\nonumber\\
&\dfn&\sup_{\zeta\ge
0}\frac{1}{n}\sum_{i=0}^{n-1}
D(P_i,\sigma_i^2,\tsigma_i^2,\lambda,\zeta)-\zeta R,
\end{eqnarray}
where we have defined
\begin{eqnarray}
D(P,\sigma^2,\tsigma^2,\lambda,\zeta)&=&\min_{\rho,\sigma_Y^2}\left\{
\frac{1}{2}\ln(2\pi\sigma^2)+\frac{\sigma_Y^2-2\sqrt{P}\sigma_Y\rho+P}{2\sigma^2}
-\frac{1}{2}\ln[2\pi
e\sigma_Y^2(1-\rho^2)]+\right.\nonumber\\
& &\left.\frac{\zeta}{\lambda}
\left[\frac{\sqrt{P}\sigma_Y\rho}{\tsigma^2}-W\left(\frac{\sqrt{P}
\sigma_Y}{\tsigma^2},\lambda\right)\right]\right\}.
\end{eqnarray}
After the minimization over $\lambda\ge 0$, and after taking the limit of
$n\to\infty$, we obtain
\begin{eqnarray}
\varepsilon(R)&=&\sup_{\zeta\ge 0}\left[\sup_{\lambda\ge
0}\frac{1}{2\pi}\int_{0}^{2\pi}
D[S_X(\omega),S_Z(\omega),\tilde{S}_Z(\omega),\lambda,\zeta]\mbox{d}\omega-\zeta
R\right]\nonumber\\
&\dfn&\sup_{\zeta\ge 0}[\Delta(\zeta)-\zeta R],
\end{eqnarray}
and it is not difficult to check that $\varepsilon(0)=2E_{\mbox{\tiny trc}}^-(0)$, 
as for $R=0$, the optimum $\lambda$ tends to infinity, in which case,
$W(\sqrt{P}\sigma_Y/\tsigma^2,\lambda)$ vanishes.
Thus $R_{\mbox{\tiny t}}$, 
the maximum guaranteed rate of tightness of our TRC bound, is the
supremum of all rates $R$ such that $\varepsilon(R)> E_{\mbox{\tiny trc}}(R)$.
To find an expression for $R_*$, we require that
for every $\theta\ge 1$, there exists $\zeta > 2\theta-1$ such that
$$\Delta(\zeta)-\zeta R\ge B(\theta)-(2\theta-1)R,$$
or equivalently,
$$R\le \frac{\Delta(\zeta)-B(\theta)}{\zeta-2\theta+1}.$$
Therefore,
\begin{equation}
R_{\mbox{\tiny t}}=\inf_{\theta\ge 1}
\sup_{\zeta>2\theta-1}\frac{\Delta(\zeta)-B(\theta)}{\zeta-2\theta+1}.
\end{equation}

\end{document}